\documentclass[a4paper,12pt]{article}
\usepackage{amsfonts,amsthm,amsmath,amssymb,graphicx}
\usepackage{color}
\begin{document}

\newtheorem{definition}{Definition}[section]
\newcommand{\be}{\begin{equation}}
\newcommand{\ee}{\end{equation}}
\newcommand{\bea}{\begin{eqnarray}}
\newcommand{\eea}{\end{eqnarray}}
\newcommand{\LE}{\left[}
\newcommand{\R}{\right]}
\newcommand{\nn}{\nonumber}
\newcommand{\Tr}{\text{Tr}}
\newcommand{\N}{\mathcal{N}}
\newcommand{\G}{\Gamma}
\newcommand{\vf}{\varphi}
\newcommand{\LL}{\mathcal{L}}
\newcommand{\Op}{\mathcal{O}}
\newcommand{\HH}{\mathcal{H}}
\newcommand{\arctanh}{\text{arctanh}}
\newcommand{\up}{\uparrow}
\newcommand{\down}{\downarrow}
\newcommand{\rd}{\partial}
\newcommand{\de}{\partial}
\newcommand{\ba}{\begin{eqnarray}}
\newcommand{\ea}{\end{eqnarray}}
\newcommand{\db}{\bar{\partial}}
\newcommand{\we}{\wedge}
\newcommand{\ca}{\mathcal}
\newcommand{\lr}{\leftrightarrow}
\newcommand{\f}{\frac}
\newcommand{\s}{\sqrt}
\newcommand{\vp}{\varphi}
\newcommand{\hvp}{\hat{\varphi}}
\newcommand{\tvp}{\tilde{\varphi}}
\newcommand{\tp}{\tilde{\phi}}
\newcommand{\ti}{\tilde}
\newcommand{\ap}{\alpha}
\newcommand{\pr}{\propto}
\newcommand{\mb}{\mathbf}
\newcommand{\ddd}{\cdot\cdot\cdot}
\newcommand{\no}{\nonumber \\}
\newcommand{\la}{\langle}
\newcommand{\lb}{\rangle}
\newcommand{\ep}{\epsilon}
 \def\we{\wedge}
 \def\lr{\leftrightarrow}
 \def\f {\frac}
 \def\ti{\tilde}
 \def\ap{\alpha}
 \def\pr{\propto}
 \def\mb{\mathbf}
 \def\ddd{\cdot\cdot\cdot}
 \def\no{\nonumber \\}
 \def\la{\langle}
 \def\lb{\rangle}
 \def\ep{\epsilon}
\def\m{{\mu}}
 \def\w{{\omega}}
 \def\n{{\nu}}
 \def\ep{{\epsilon}}
 \def\d{{\delta}}
 \def\rh{\rho}
 \def\t{{\theta}}
 \def\a{{\alpha}}
 \def\T{{\Theta}}
 \def\frac#1#2{{#1\over #2}}
 \def\l{{\lambda}}
 \def\G{{\Gamma}}
 \def\D{{\Delta}}
 \def\g{{\gamma}}
 \def\s{\sqrt}
 \def\ch{{\chi}}
 \def\b{{\beta}}
 \def\CA{{\cal A}}
 \def\CC{{\cal C}}
 \def\CI{{\cal I}}
 \def\CO{{\cal O}}
 \def\o{{\rm ord}}
 \def\Ph{{\Phi }}
 \def\L{{\Lambda}}
 \def\CN{{\cal N}}
 \def\p{\partial}
 \def\pslash{\p \llap{/}}
 \def\Dslash{D \llap{/}}
 \def\Mp{m_{{\rm P}}}
 \def\apm{{\alpha'}}
 \def\r{\rightarrow}
 \def\Re{{\rm Re}}
 \def\MG{{\bf MG:}}
\def\be{\begin{equation}}
\def\ee{\end{equation}}
\def\ba{\begin{eqnarray}}
\def\ea{\end{eqnarray}}
\def\bal{\begin{align}}
\def\eal{\end{align}}

 \def\de{\partial}
 \def\db{\bar{\partial}}
 \def\we{\wedge}
 \def\lr{\leftrightarrow}
 \def\f {\frac}
 \def\ti{\tilde}
 \def\ap{\alpha}
  \def\al{\alpha'}
 \def\pr{\propto}
 \def\mb{\mathbf}
 \def\ddd{\cdot\cdot\cdot}
 \def\no{\nonumber \\}
\def\nn{\nonumber \\}
 \def\la{\langle}
 \def\lb{\rangle}
 \def\ep{\epsilon}
 \def\ddbp{\mbox{D}p-\overline{\mbox{D}p}}
 \def\ddbt{\mbox{D}2-\overline{\mbox{D}2}}
 \def\ov{\overline}
 \def\cl{\centerline}
 \def\vp{\varphi}
\def\hB{\hat \Box}

\begin{titlepage}

\thispagestyle{empty}

\begin{flushright}
YITP-18-93\\
IPMU18-0139\\
\end{flushright}

\vspace{.4cm}
\begin{center}
\noindent{\large \bf
Holographic Spacetimes as Quantum Circuits of Path-Integrations}\\
\vspace{2cm}

Tadashi Takayanagi$^{a,b}$
\vspace{1cm}

{\it
$^{a}$Center for Gravitational Physics, \\
Yukawa Institute for Theoretical Physics (YITP), Kyoto University, \\
Kitashirakawa Oiwakecho, Sakyo-ku, Kyoto 606-8502, Japan\\ \vspace{3mm}
$^{b}$Kavli Institute for the Physics and Mathematics of the Universe,\\
University of Tokyo, Kashiwano-ha, Kashiwa, Chiba 277-8582, Japan\\
}

\vskip 2em
\end{center}

\vspace{.5cm}
\begin{abstract}
We propose that holographic spacetimes can be regarded as collections of quantum circuits
based on path-integrals. We relate a codimension one surface in a gravity dual to a quantum circuit given by a path-integration on that surface with an appropriate UV cut off. Our proposal naturally generalizes the conjectured duality between the AdS/CFT and tensor networks. This largely strengthens the surface/state duality and also
provides a holographic explanation of path-integral optimizations.
For static gravity duals, our new framework provides a derivation of the holographic complexity formula given by the gravity action on the WDW patch. We also propose a new formula which relates numbers of quantum gates to  surface areas, even including time-like surfaces, as a generalization of the holographic entanglement entropy formula. We argue the time component of the metric in AdS emerges from the density of unitary quantum gates in the dual CFT. Our proposal also provides a heuristic understanding how the gravitational force emerges from quantum circuits.
\end{abstract}

\end{titlepage}
\tableofcontents
\newpage

\section{Introduction}

The idea of holography has changed our standard notion of spacetime in the presence of gravitational force \cite{Hol}. The AdS/CFT provides us with ideal setups to study holography in a microscopic way \cite{Ma,GKP,Witten}. The considerations of holographic entanglement entropy
reveal deep connections between gravity and quantum information \cite{RT,HRT,ReviewHEE}.
In particular, this suggests that gravitational spacetimes may emerge from geometric structures of
quantum entanglement in conformal field theories (CFTs)
or more generally quantum many-body systems. One concrete idea to realize this emergent
spacetime is to employ tensor networks as first conjectured in \cite{TNa}.
For other interesting approaches to emergent spacetimes from quantum entanglement, refer to e.g. \cite{VRa,MaSu,FH}.
It was also argued that quantum error correcting codes may also provide another explanation of
the emergent bulk spacetimes \cite{ADH}.

The tensor network is a graphical method to describe a quantum many-body wave function in terms of a network
of quantum entanglement (see e.g. the reviews \cite{CiVe,TG}). In the original conjecture \cite{TNa}, it was argued that a canonical time slice (i.e. a hyperbolic space) in an AdS corresponds to a special tensor network called MERA (multi-scale entanglement renormalization ansatz) \cite{MERA}. The MERA gives useful tensor networks which produce CFT vacua. Its continuous version called cMERA was also defined in \cite{cMERA,Cotler} and applied to the AdS/CFT \cite{NRT}. Later, a modified correspondence which argues that the MERA corresponds to a de-Sitter space was proposed \cite{Beny,Cz,Car} based on the causal structure of MERA. On the other hand, a tensor network, called the perfect tensor network, was introduced which is expected to describe a hyperbolic disk, based on the quantum error correcting codes \cite{TNc}. Its refined version called random tensor networks was also
constructed \cite{TNd} and its spacetime version was formulated in \cite{QY}.
However in these models, which is different from the MERA, the resulting states typically deviate from CFT vacua.

There is another approach which starts from Euclidean path-integral description of the CFT vacuum and which employs a procedure called the path-integral optimization \cite{MTW,Caputa:2017urj} (refer also to \cite{CzechC,Bhattacharyya:2018wym,Molina-Vilaplana:2018sfn} for later developments). This reproduces the correct metrics of canonical time slices after the optimization. This approach was motivated by a tensor network picture of AdS/CFT because
we can regard a discretized version of Euclidean path-integrals as certain tensor networks, which are not necessarily isometric. Explicit relations between the tensor networks and path-integrations have recently been worked out in \cite{TNR,Path}. Nevertheless, so far it has not been fully clear how and why the path-integrations should be embedded in the full AdS geometry.

\begin{figure}
  \centering
 \includegraphics[width=5cm]{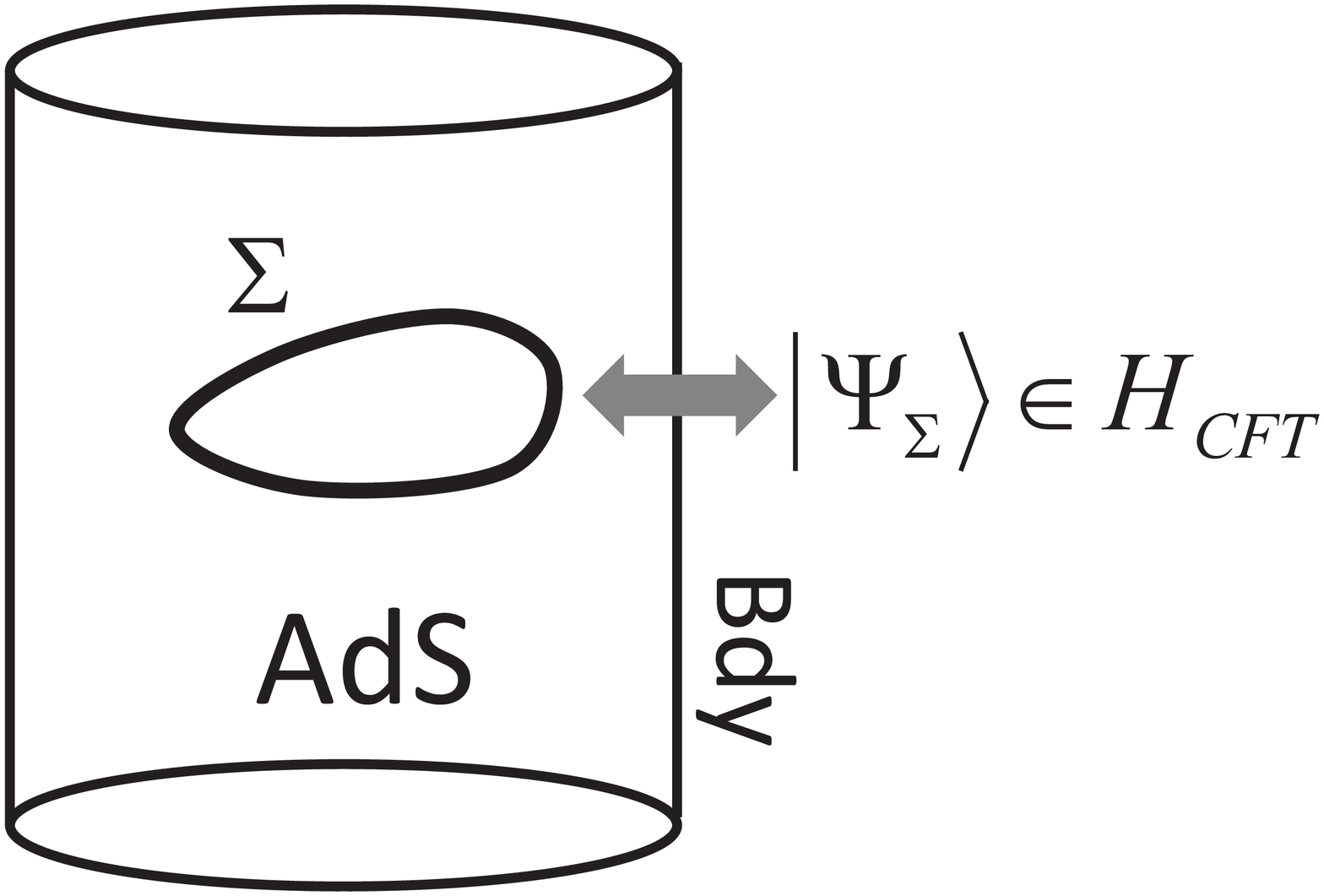}
  \includegraphics[width=5cm]{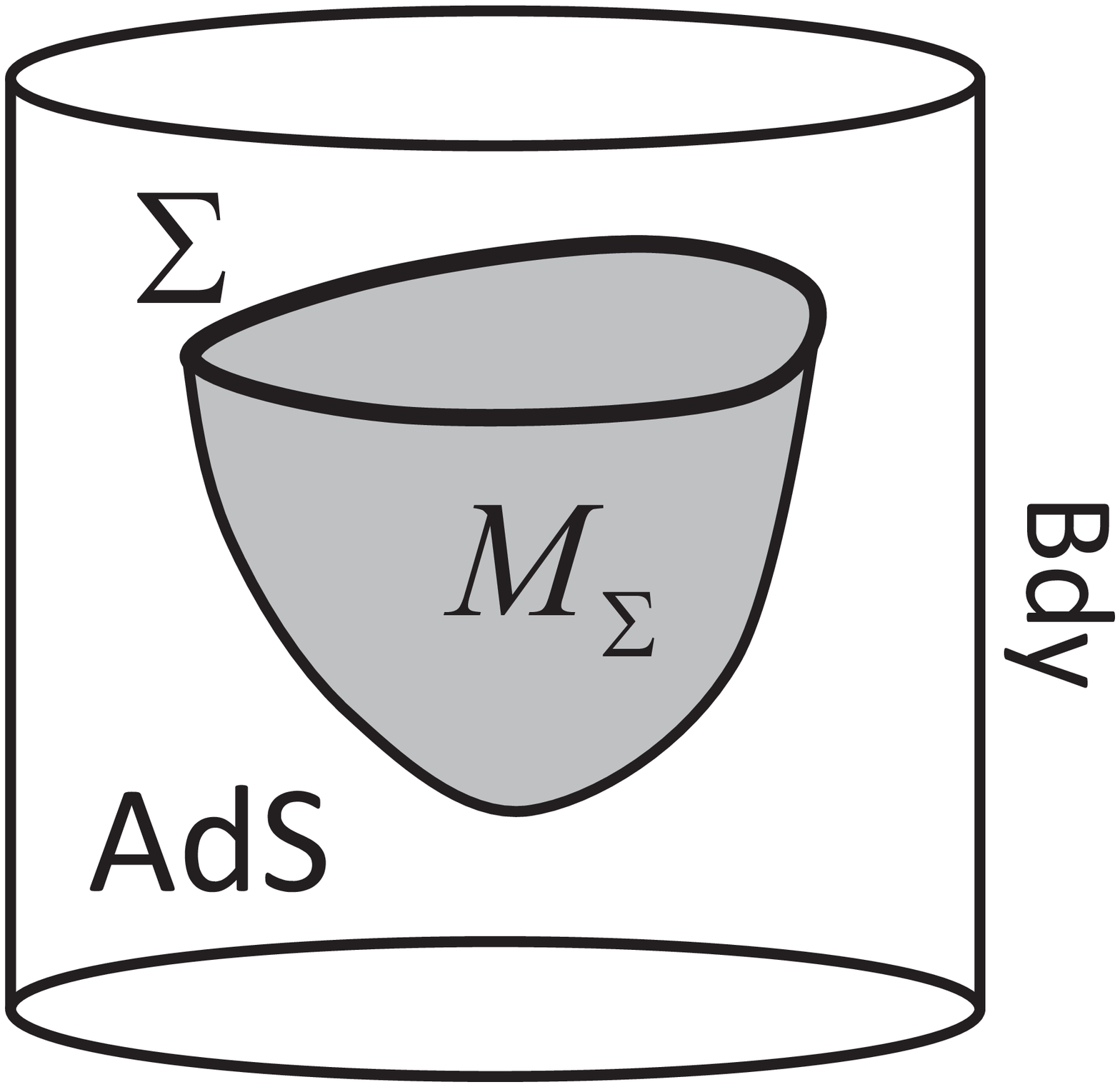}
 \caption{The left picture is a sketch of surface/state correspondence in the context of AdS/CFT \cite{MiTa}. The right picture explains the new correspondence proposed in the present paper, based on path-integrations in surface/state duality for Euclidean AdS.}
\label{fig:SSduality}
\end{figure}

In this way, even though there have been remarkable developments on connections between
AdS/CFT and tensor networks, we still do not know precisely which tensor network corresponds to which
surface in AdS. Especially we do not understand well how to interpret the time component of the metric $g_{tt}$ in a gravity dual from the tensor networks. In such a situation, it is helpful to study things in an opposite way: we start with a holographic spacetime of a gravity dual and look at its surfaces to ask what they correspond to in tensor networks.
A partial step has been taken in our earlier work \cite{MiTa} (see \cite{MMiyaji,cSS} for related progresses), where the surface/state correspondence was proposed. Refer to \cite{NomuraA,NomuraB,NomuraC} for interesting works on similar problems. See also \cite{TTbar} for another intriguing proposal to move the AdS boundary in the bulk.
This surface/state correspondence argues that an arbitrary convex codimension two surface $\Sigma$ in AdS corresponds to a certain quantum state $|\Psi_\Sigma\lb$ in the dual CFT as in the left picture in Fig.\ref{fig:SSduality}.

In the present article, we would like to combine the above two ideas: path-integral optimization and surface/state correspondence. We propose a new framework of holography where each codimension one surface $M_\Sigma$ in the gravity dual is interpreted as a quantum circuit defined by a path-integration on $M_\Sigma$ with a suitable UV cut off,
both in Lorentzian and Euclidean signature. Refer to the right picture in Fig.\ref{fig:SSduality}. Here we discretize path-integrations of CFTs into those on lattices and regard them as quantum circuits.
Our proposal largely generalizes and clarifies the conjectured correspondence between tensor networks and slices in AdS in a covariant way including the time coordinate. Note that if $M_\Sigma$ is located in the AdS boundary, our proposal just follows from the standard bulk-boundary correspondence in AdS/CFT \cite{GKP,Witten}.

The other motivation of this paper is to understand the holographic calculations of complexity.
Recently the computational complexity for quantum states in CFTs has been studied actively because
it may provide a new window which connects gravity to quantum information theory \cite{SUR,Susskind}. In particular,
a holographic formula which computes the complexity was proposed in \cite{Comp} (see \cite{Lehner:2016vdi} for corner contributions), where the complexity is given by the gravity action restricted to a region called a Wheeler DeWitt (WDW) patch. For recent developments of holographic complexity, refer to e.g. \cite{MNSTW,Ali,BaRa,subreg2,CMM,Simon,Growth,Kim,Zhaoy,SW,Maloney,
Susskind:2018tei,BRR,Chenc,Ageev:2018nye,AHS,HISS,MyV}.

However, there is no clear derivation of this formula even if we assume the AdS/CFT correspondence.
This is partly because the definition of computational complexity is so involved in quantum field theories (QFTs) that no unique calculational scheme was established so far, as opposed to the calculations of entanglement entropy.
Nevertheless, explicit calculations of complexity in QFTs have been performed based on plausible definitions \cite{CHMP, JeMy} and interesting results have been obtained (refer to e.g.
\cite{HIS,KKS,HaMy,Magan,AlCa,CaMa,CCDHJ,GHMR,BSS}). Also in the framework of the path-integral optimization \cite{Caputa:2017urj} for two dimensional CFTs, the complexity functional is identified with the Liouville action.
This provides a `field theory friendly' approach and is called the path-integral complexity. An interesting connection between the original definition of complexity in terms of quantum circuits and the Liouville action has been uncovered quite recently in \cite{CaMa}.

In our new framework using the path-integral circuits, we generalize the holographic correspondence and define a quantity called holographic path-integral complexity in gravity duals. Interestingly, for static Lorentzian setups, we manage to show that the holographic path-integral complexity, which has a clear definition in dual CFTs, essentially agrees with the holographic complexity based on the gravity action in the WDW patch. For time-dependent quantum states, our holographic path-integral complexity prescription does not seem to precisely coincide with the previous holographic complexity. We will also introduce a new connection between quantum entanglement and geometry,
which argues that an area element in a gravity dual can be interpreted as the maximal amount of entanglement entropy created by the corresponding quantum gates. These new relations will enable us to conclude that the time component of the metric in AdS emerges from the density of unitary quantum gates which scramble quantum states in the dual CFT.

This paper is organized as follows: In section two, we will describe our new framework of holographic correspondence between codimension one surfaces and quantum circuits described by path-integrations. In section three, we define and evaluate the holographic path-integral complexity in our framework. In section four, we study the evolution of quantum entanglement under the quantum circuits of path-integrations and propose a formula which relates an area of
surface and the number of quantum gates which add quantum entanglement. In section five, we summarize our conclusions and discuss future problems. In appendix A, we gave a derivation of Liouville action from the gravity action for AdS$_3$. In appendix B, we analyzed examples of quantum circuits of path-integrals in two dimensional CFTs, which corresponds to a de Sitter space and hyperbolic space.

\section{AdS as Quantum Circuits of Path-Integrations}

The surface/state duality \cite{MiTa} argues that an arbitrary $d$ dimensional (i.e. codimention two) connected closed surface $\Sigma$ which is convex and space-like in a $d+2$ dimensional gravitational spacetime $N_{d+2}$ (either Euclidean or Lorentzian), corresponds to a certain quantum state $|\Psi_\Sigma\lb$ in a Hilbert space ${\cal H}_{N}$ specific to the spacetime $N_{d+2}$:
\be
\Sigma_{d}\in N_{d+2}\lr |\Psi_\Sigma\lb \in {\cal H}_{N}.
\ee
In particular, for the AdS/CFT,  $\Sigma$ is a convex $d$ dimensional closed surface in AdS$_{d+2}$ and  ${\cal H}_N$ is identified with the CFT Hilbert space ${\cal H}_{CFT}$. Refer to the left picture in Fig.\ref{fig:SSduality}.

Below, we consider the surface/state duality in the AdS/CFT case and would like to argue that it leads to an interpretation of codimension one surfaces\footnote{
We may need to impose an analogue of convexity condition on $M_\Sigma$ as we did for the surface $\Sigma$.
We will not get into details of this issue as it does not affect our conclusions. However, one natural
constraint will be such that $M_\Sigma$ should be foliated by convex surfaces. We would like to thank
Masamichi Miyaji for discussions on this point.}  in AdS, called $M_\Sigma$, as quantum circuits of path-integrals (see the right picture in Fig.\ref{fig:SSduality}). Originally the surface/state correspondence \cite{MiTa} is motivated by a conjectured tensor network description of AdS/CFT. Here we would like to study how we can construct the state $|\Psi_\Sigma\lb$ in a CFT. We consider both Euclidean AdS and Lorentzian AdS separately below.
We will also allow a generalization of codimension two surfaces such that $\Sigma$ consists of multiple
disconnected surfaces, where the dual state $|\Psi_\Sigma\lb$ cannot be accommodated in
${\cal H}_{CFT}$, but can be included in its multiple copies.

We would like to stress that in this paper, we are focusing on a classical gravity limit of AdS/CFT correspondence, ignoring quantum fluctuations as assumed in the surface/state duality. Therefore, there is a definite AdS spacetime for a given boundary and
we consider the surface $\Sigma$ and $M_\Sigma$ in this fixed AdS spacetime.

\subsection{Euclidean AdS}

Let us first start with asymptotically AdS backgrounds with the Euclidean signature, which is simpler than the Lorentzian case. Indeed, the surface/state correspondence \cite{MiTa} was originally proposed for Euclidean spaces. Our main claim in the present article is that each state $|\Psi_\Sigma\lb$  can be obtained from a regularized path-integration on a codimension one surface $M_\Sigma$, which ends on the surface $\Sigma$ i.e.
$\de M_{\Sigma}=\Sigma$. Refer to the right picture in Fig.\ref{fig:SSduality}.
Note that the choice of $M_{\Sigma}$ is not unique and indeed there are infinitely many different surfaces which satisfy the condition $\de M_{\Sigma}=\Sigma$.
 Our claim is summarized as
\ba
e^{C(M_\Sigma)}\cdot \Psi_\Sigma[\vp_0(x)]=\int \left[\prod_{y\in M_\Sigma}D\vp(y)\right]
e^{-S^{CFT}_{M_{\Sigma}}[\vp]}\prod_{x\in \Sigma}\delta(\vp(x)-\vp_0(x)),  \label{dual}
\ea
where we expressed all fields by the symbol $\vp$ and the action $S^{CFT}_{M_{\Sigma}}[\vp]$ for the path-integration is
the CFT action defined on $M_\Sigma$ with an appropriate regularization. The wave functional
$\Psi_\Sigma[\vp_0(x)]$ is normalized such that it has a unit norm, where overall normalization contributions are expressed as the factor $e^{C(M_\Sigma)}$.
In the coordinate system of the Poincare AdS$_{d+2}$ with the radius $R_{AdS}$:
\ba
ds^2=R^2_{AdS}\left(\frac{dz^2+dt^2+\sum_{i=1}^d dx_i^2}{z^2}\right),  \label{pads}
\ea
the regularization is such that the lattice spacing is given by $z$. More generally the lattice regularization of $S^{CFT}_{M_{\Sigma}}[\vp]$ should be done such that one lattice site corresponds to the unit area measured by the dimension less metric $ds^2/R^2_{AdS}$. The constant $C(M_\Sigma)$ is called the path-integral complexity, which is essentially the same one introduced in \cite{Caputa:2017urj}.

\begin{figure}
  \centering
  \includegraphics[width=7cm]{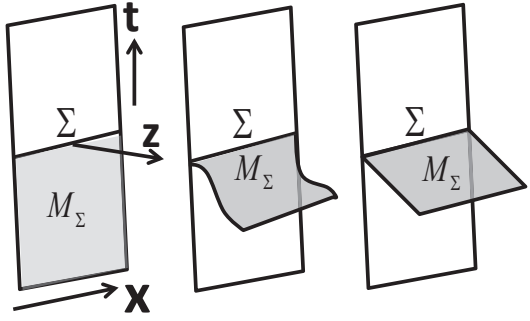}
 \caption{Various constructions of the same state $|\Psi_\Sigma\lb$ (i.e. the vacuum state in a CFT) from path-integrations on different surfaces $M_\Sigma$ and their gravity duals. \label{fig:SSpathpath}}
\end{figure}

Now, as the simplest example, consider a pure Euclidean AdS setup.
We take $\Sigma$ to be the time slice at the AdS boundary i.e. $z=\ep$ and $t=0$ in the coordinate
(\ref{pads}), which is depicted as the left picture in Fig.\ref{fig:SSpathpath}.
In this case the state $|\Psi_\Sigma\lb$ coincides with the CFT vacum $|0\lb$ with the lattice constant given by $\ep$. If we choose $M_\Sigma$ to be the path-integral along the time coordinate $t$ in Euclidean AdS, the conjectured formula (\ref{dual}) coincides with the standard Eulidean path-integral which produces the ground state wave functional.

If we choose a generic\footnote{
In this paper we assume that the surface $M_\Sigma$ has the simplest topology namely the disk in this setup. We expect this restriction comes from the known convexity and topology condition in surface/state duality, though we would like to leave the detail for future works.} $d+1$ dimensional
surfaces as $M_\Sigma$ as in the middle picture in Fig.\ref{fig:SSpathpath}, one may worry that the state $|\Psi_\Sigma\lb$ in  (\ref{dual}) depends not only on $\Sigma$ but also on the choice of $M_\Sigma$. However, we would like to argue the state obtained after the path-integration on $M_\Sigma$ does not depend on the choice of $M_\Sigma$ owing to the conformal invariance of CFT. In the setup of AdS$_3/$CFT$_2$, this independence on the choice of $M_\Sigma$ is obvious because metrics of two surfaces $M^{(1)}_\Sigma$ and $M^{(2)}_\Sigma$ with the same topology are related by a Weyl transformation.

We would like to suggest that the same property can be true also in the higher dimensional 
AdS/CFT. First let us note that
the Weyl transformation has degrees of freedom of one function $\phi(x,z)$ on a $d+1$ dimensional space: $g_{ab}\to e^{2\phi(x,z)}g_{ab}$. This agrees with the degrees of 
freedom of the choice of $M_\Sigma$ which is specified by the time slice $t=t(x,z)$. 
Consider the following form of the AdS metric
\be
\frac{ds^2}{R^2_{AdS}}=d\rho^2+\cosh^2\rho\left(\frac{dt^2+dy^2+d\eta^2}{\eta^2}\right).
\ee
If we choose $M_\Sigma$ to be the surface $\rho=\rho(t,y,\eta)$, then near the AdS boundary $\rho\to \infty$
we have
\be
\frac{ds^2}{R^2_{AdS}}\simeq \cosh^2\rho\left(\frac{dt^2+dy^2+d\eta^2}{\eta^2}\right),
\ee
which is indeed the conformal transformation of the flat spacetime where the CFT vacuum $|0\lb$ was defined by the path-integral. In this way, we can relate two different 
surfaces $M^{(1)}_\Sigma$ and $M^{(2)}_\Sigma$ by a Weyl transformation in the UV region of AdS. Apparently, this relation is lost once we consider the full AdS space including the 
IR regions, except special examples, such as the case where $M_\Sigma$ is given by a hyperbolic space. However, in the IR regions, we cannot apply the usual notion of Weyl invariance in the continuum limit and will need other treatments whose details are beyond the scope of this paper.

This procedure of increasing the coarse-grainings without
changing the final quantum state $|\Psi_\Sigma\lb$ corresponds to the path-integral optimization
introduced in \cite{Caputa:2017urj}. This optimization eventually leads to the hyperbolic surface
 $t=0$, as depicted in the right picture of Fig.\ref{fig:SSpathpath}, and is expected to be the most efficient Euclidean path-integration to produce the CFT vacuum.

On the other hand, if we move the vertical surface $z=\ep$ toward the inside of AdS to $z=z_0$, then we expect that the path-integration along the time direction can be done by employing an action, which is
coarse-grained up to the length-scale $z_0$. This correspondence between 
the radial coordinate of AdS and the effective cut off scale (or equally coarse-graining)
is given manifestly in the formulation of path-integral optimization which will be 
discussed in the next subsection. In addition, we can deform the shape of such a surface. In this way we can interpret surfaces in an Euclidean AdS as (non-unitary) quantum circuits of Euclidean path-integrations with an appropriate UV cut off.

Moreover, we expect that the above argument using the Weyl invariance for the pure AdS
can also be applied to general asymptotically AdS backgrounds by considering the relevant perturbations of holographic CFTs as in the massive path-integral optimization, done recently in \cite{Bhattacharyya:2018wym}.

\subsection{Relation to Path-integral Optimization}\label{pio}

The invariance of quantum states under Weyl transformations of Euclidean path-integrations
has been recently employed in \cite{Caputa:2017urj,Bhattacharyya:2018wym} to optimize the path-integral computations. For two dimensional (2d) CFTs, we can write the metric on the space $M_\Sigma$ where we perform the path-integration in the form
\be
ds^2_{CFT}=e^{2\phi(t,x)}(dt^2+dx^2),  \label{cftm}
\ee
where $t$ is the Euclidean time. The rule of UV regularization is such that one lattice site corresponds to a unit area in the above metric. The surface $\Sigma$ is specified by $t=-\ep$ and the path-integration is performed for $-\infty<t<-\ep$.

The path-integral complexity is given by
the Liouville action
\be
C_L(M_\Sigma)=\frac{c}{24\pi}\int^{-\ep}_{-\infty} dt\int dx \left[(\de_t\phi)^2+(\de_x\phi)^2+e^{2\phi} \right],  \label{lvlacet}
\ee
where $c$ is the central charge of the 2d CFT. Indeed as we show in appendix A, we can derive the Liouville action from the bulk AdS$_3$ action with the boundary metric (\ref{cftm}).

The optimization is performed by minimizing $C_L(M_\Sigma)$ with respect to $\phi(t,x)$ with the boundary condition $e^{\phi(t=-\ep,x)}=1/\ep$, which ensures
that we obtain the expected quantum state with the UV regularization scale $\ep$.
This leads to the solution
\be
e^{\phi(t,x)}=\frac{1}{|t|},
\ee
and thus the space $M_\Sigma$ is given by the hyperbolic space. We can identify this optimized surface $M_\Sigma$ with the $t=0$ time slice of the Poincare AdS$_3$ (\ref{pads}), where we identify the AdS
metric $ds^2/R^2_{AdS}$ on $M_\Sigma$ with the CFT metric $ds^2_{CFT}$ (\ref{cftm}). Indeed the processs of modifying the space of Euclidean path-integrations corresponds to the change of surface $M_\Sigma$ as in Fig.\ref{fig:SSpathpath}.

Note that though our correspondence (\ref{dual}) works for any surface $M_\Sigma$, the path-integral optimization picks up a special surface which minimizes $C_L(M_\Sigma)$. For a static asymptotically AdS space, the minimization chooses the canonical time slice with the minimal volume. Thus we naturally understand the observation found in  \cite{Caputa:2017urj,Bhattacharyya:2018wym} that an optimized metric agrees with the metric on the constant time slice in its gravity dual (refer to the right picture in Fig.\ref{fig:SSpathpath}).

\subsection{Lorentzian AdS}

\begin{figure}
  \centering
  \includegraphics[width=7cm]{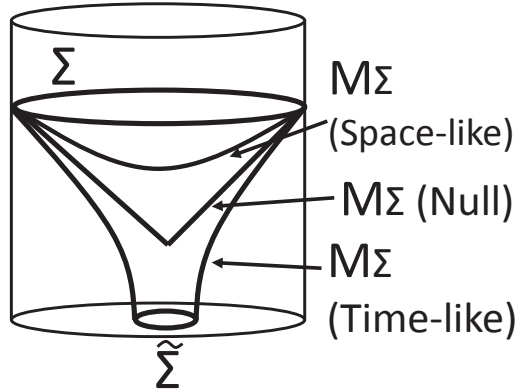}
 \caption{The path-integral construction of the state $|\Psi_\Sigma\lb$ when we take $\Sigma$ in Lorentzian AdS. $M_\Sigma$ can be either time-like, null or space-like.}
\label{fig:SSpathtime}
\end{figure}

It is quite natural to expect that the surface/state correspondence is true also for Lorentzian AdS.
However, the situation is a little more complicated because in this case, the surface $M_{\Sigma}$ can be either time-like, null or space-like as in Fig.\ref{fig:SSpathtime}. We would like to conjecture that
when $M_{\Sigma}$ is time-like, the state $|\Psi_\Sigma\lb$ is obtained by a Lorentzian path-integral on $M_\Sigma$ with an appropriate cut off as a simple extension of our conjecture in the Euclidean setup.
In other words, we have
\ba
e^{iC(M_\Sigma)}\cdot \Psi_\Sigma[\vp_0(x)]=\int \left[\prod_{y\in M_\Sigma}D\vp(y)\right]
e^{iS^{CFT}_{M_{\Sigma}}[\vp]}\prod_{x\in \Sigma}\delta(\vp(x)-\vp_0(x)).  \label{Tdual}
\ea
We call $C(M_\Sigma)$ as the path-integral complexity in the Lorentzian case. 
This framework allows us to interpret
time-like surfaces $M_\Sigma$  as quantum circuits. Moreover, if $M_\Sigma$ is null, we can understand it as a degenerate limit of time-like surfaces. Note that the appearance of the phase factor $e^{iC(M_\Sigma)}$ is
consistent with the form of gravity partition function $e^{iI_G}$ in the Lorentzian signature.

When $M_{\Sigma}$ is space-like, we would like to argue that basically it corresponds to
a path-integral on the space-like surface $M_{\Sigma}$.
One may worry that the Euclidean path-integration changes the
normalization of wave functional and this might contradict with the Lorentzian evolution of the gravity dual.
However as we will see in section \ref{Lorentzian}, a careful analysis of corner contributions in the gravity dual shows the presence of such a change of normalization. Also notice that if there is a purely unitary (=Lorentzian) quantum circuit on a space-like surface $M_\Sigma$, then propagations of local excitations can break the causality in the bulk AdS. Therefore the circuit should be non-unitary. As our later result of path-integral complexity imply, we expect the quantum circuit on $M_{\Sigma}$ includes not only non-unitary but also unitary quantum gates.

As the simplest example, consider a Lorentzian pure AdS. In this case, the state $|\Psi_\Sigma\lb$ coincides with the CFT vacuum state $|0\lb$.  The time-like path-integration starts with another quantum state $|\Psi_{\ti{\Sigma}}\lb$ dual to the surface $\ti{\Sigma}$ (see Fig.\ref{fig:SSpathtime}).
We can identify $|\Psi_{\ti{\Sigma}}\lb$ with the vacuum state $|0\lb_{\ti{\Sigma}}$, with the coarse-graining specified by the surface $\ti{\Sigma}$. It is clear that the path-integration on the
time-like surface does not affect modes whose wave lengthes are larger than the ones in $\ti{\Sigma}$, owing to the Weyl invariance. On the other hand, this time-like path-integration creates vacuum state for the modes whose wave lengthes are between the one for $\ti{\Sigma}$ and the one for $\Sigma$.
The Weyl invariance of path-integrations explains that for any choice of the codimension one time-like surface $M_\Sigma$ which connects $\Sigma$ and $\ti{\Sigma}$, its dual quantum circuit maps
$|\Psi_{\ti{\Sigma}}\lb$ into $|\Psi_\Sigma\lb$.

\section{Holographic Path-Integral Complexity in AdS}\label{PathC}

In the previous section, we argued that the codimension one surface $M_\Sigma$ in AdS
can be regarded as a path-integration with a suitable cut off. By introducing a discretization with the cut off scale, this path-integration can also be regarded as a quantum circuit. In this section we would like to consider how this quantum circuit generates a computational complexity. The computational complexity is originally defined as the number of quantum gates, whose precise definition in field theories involves subtleties and is not completely understood at present. Instead, we consider a quantity called the path-integral complexity $C(M_\Sigma)$ \cite{Caputa:2017urj} defined in (\ref{dual}), whose definition in field theories is straightforward. This quantity measures the size of path-integration and therefore is expected to be proportional to the complexity. We will analyze the
holographic counterpart of path-integral complexity and compare our results with the earlier holographic complexity proposal in \cite{Comp}.
We will study both asymptotically Euclidean and Lorentzian AdS setups below separately.

\subsection{Holographic Path-Integral Complexity in Euclidean AdS}\label{Euclidean}

First, we focus on the cases where gravity duals are given by asymptotically Euclidean
AdS spaces with static metrics. We choose $\Sigma$ such that it is a codimension two convex surface on a canonical time slice $t=0$.

We can compute the path-integral complexity $C(M_\Sigma)$ by employing the obvious relation
\be
e^{2C(M_\Sigma)}=\la \Psi_\Sigma(M_\Sigma)|\Psi_\Sigma(M_\Sigma)\lb, \label{ppp}
\ee
where we write the state $|\Psi_\Sigma\lb$ as $|\Psi_\Sigma(M_\Sigma)\lb$ by emphasizing that we performed the path-integration on $M_\Sigma$. Notice that the states $|\Psi_\Sigma(M_\Sigma)\lb$ for various choices of $M_\Sigma$ are the same state, denoted by $|\Psi_\Sigma\lb$ as before, up to the overall normalization, which is proportional to $e^{C(M_\Sigma)}$.

\begin{figure}[tttt]
  \centering
  \includegraphics[width=8cm]{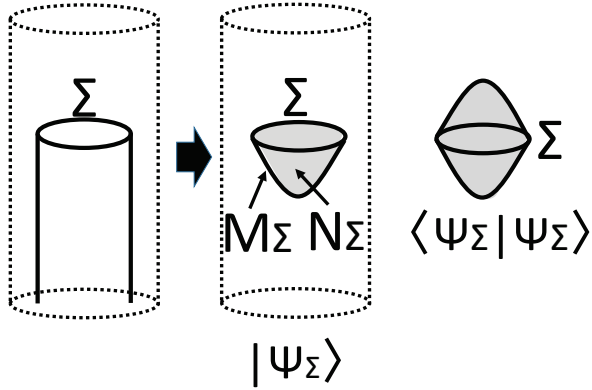}
 \caption{The sketch of holographic computation of path-integral complexity in a global Euclidean AdS.
 For simplicity, we choose $\Sigma$ is a codimension two convex surface on a time slice $t=0$.
 Originally, the state $|\Psi_\Sigma\lb$ dual to the surface $\Sigma$ is obtained by the path-integration along the Euclidean time with a coarse-grained CFT action as in the left picture. Then we can deform the space $M_\Sigma$ on which we perform the path-integration without changing the quantum state $|\Psi_\Sigma\lb$ as depicted in the middle picture. The bulk region surrounded by the time slice $t=0$ and $M_{\Sigma}$ is called $N_\Sigma$.
 During this process we can reduce the normalization of
 wave function and this normalization is computed by doubling the system namely the inner product
 $\la\Psi_\Sigma|\Psi_{\Sigma}\lb$. In the gravity dual, this inner product is given by the gravity action evaluated on the Euclidean space given by a double copy of $N_\Sigma$, depicted in the right picture.}
\label{fig:complexity}
\end{figure}

By extending the standard bulk-boundary relation to our finite cut off surface, we can calculate (\ref{ppp}) as the gravity partition function on $N_\Sigma$
as depicted in Fig.\ref{fig:complexity}.
The (coarse-grained) CFT on $M_\Sigma$ is dual to the bulk space $N_{\Sigma}$, which is defined by the region surrounded by the canonical time slice $t=0$  and the surface $M_\Sigma$. Then the complexity
$C(M_\Sigma)$ for the state $|\Psi_\Sigma\lb$ is computed as
\be
e^{C(M_\Sigma)}=e^{-I^E_{G}(N_\Sigma)},  \label{Ecpxs}
\ee
where $I^E_{G}(N_\Sigma)$ is the value of the total Euclidean gravity action in $N_\Sigma$ and we employed the bulk-boundary relation.

The path-integral complexity of the state $|\Psi_\Sigma\lb$, written as
$C(|\Psi_\Sigma\lb)$ is defined by taking the minimum over all possible choice of $M_\Sigma$:
\ba
C(|\Psi_\Sigma\lb)=\min_{M_\Sigma} C(M_\Sigma)=\min_{M_\Sigma}\left[-I^E_{G}(N_\Sigma)\right].
\ea

As a simple example, consider a Poincare AdS$_3$ given by the metric (\ref{pads}) and calculate the
path-integral complexity for the vacuum state, where the codimension one surface $\Sigma$ is given by the straight line $-\infty<x<\infty$
at $z=\ep$ and $t=0$. We choose the surface $M_{\Sigma}$ to be the semi-infinite line
\be
M_{\Sigma}=\{(t,x,z)|\ t=-(z-\ep)\tan\ap\leq 0,\ \ z\geq \ep, \ \ -\infty<x<\infty\},
\label{msig}
\ee
where $\ap$ is the tilting angle of $M_\Sigma$ against $t=0$ time slice.

The Euclidean gravity action in such a setup looks like
\ba
I^E_G=-\frac{1}{16\pi G_N}\int_{N_\Sigma} \s{g}(R-2\Lambda)
-\frac{1}{8\pi G_N}\int_{M_\Sigma} \s{h}K+\frac{1}{16\pi G_N}\int_{\Sigma}\s{\gamma}(2\ap-\pi),
\label{ewww}
\ea
where $K$ is the extrinsic curvature on $M_\Sigma$ and the final term, found in \cite{Hay}, arises\footnote{Note that the corner term with the angle $2\ap$ is given by
$\frac{1}{8\pi G_N}\int_{\Sigma}\s{\gamma}(2\ap-\pi)$. Here we took a half of this because
we restrict to the lower half geometry $t\leq 0$ to describe the wave functional.}
because of the non-smooth corner of $N_\Sigma$ along $\Sigma$. By plugging explicit on-shell values
$R=6\Lambda=-\frac{6}{R^2_{AdS}}$ and $K=2\sin\ap$, we finally obtain the path-integral complexity
\be
C(M_\Sigma)=\frac{cL}{12\pi\ep}\left[\tan\ap-\ap+\frac{\pi}{2}\right], \label{eww}
\ee
where $L$ is the infinite length $\int dx$; we also employed the well-known formula $c=\frac{3R_{AdS}}{2G_N}$,
between the size of AdS and the central charge \cite{BRH}.

Thus it is clear that this reaches its minimum at $\ap=0$:
\ba
C(|\Psi_\Sigma\lb)=\frac{cL}{24\ep}, \label{ewwa}
\ea
where $M_\Sigma$ coincides the hyperbolic space $H_2$ defined by the canonical time slice $t=0$.

On the other hand the maximum value of  $C(M_\Sigma)$ is achieved at $\ap=\frac{\pi}{2}$ i.e. the standard Euclidean path-integral on a flat space, where we obtain
\be
\max_{M_\Sigma}[C(M_\Sigma)]=\frac{cLT}{12\pi \ep^2},  \label{Ewwww}
\ee
where $T$ is the infinitely length in the Euclidean time $t$.

\begin{figure}
  \centering
  \includegraphics[width=8cm]{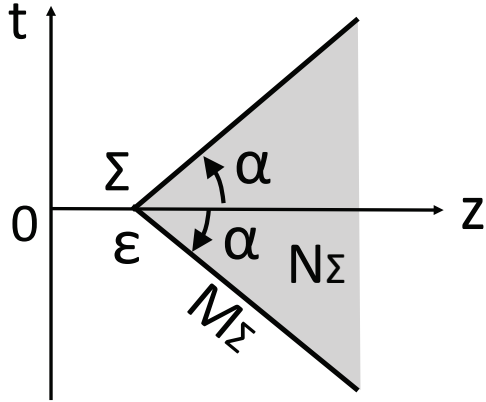}
 \caption{The setup of holographic calculation of path-integral complexity in a Poincare Euclidean AdS$_3$.}
\label{fig:complexity2}
\end{figure}

It is straightforward to extend the above computations to general static AdS/CFT setups.
By minimizing the action, it is clear that the minimum of the complexity, identified with the path-integral complexity for the quantum state $|\Psi_\Sigma\lb$, is given by the area of the corner surface $\Sigma$:
\be
C(|\Psi_\Sigma\lb)=\frac{1}{16G_N}\int_{\Sigma} \s{\gamma}=\frac{S(\Sigma)}{4},  \label{cpxent}
\ee
where
\be
S(\Sigma)=\frac{A(\Sigma)}{4G_N},
\ee
is the `entropy' for the surface $\Sigma$, which is obtained by applying the Bekenstein-Hawking
formula of black hole entropy to this surface. It is intriguing to note that though the bulk region $N_\Sigma$ vanishes to zero size at $\ap=0$,
the gravity action $I_G$ becomes non-trivial due to the corner angle term.

It is straightforward to extend this analysis to finite cut off surfaces $\Sigma$ such as $z=z_0$ or even those in more general static asymptotically AdS spaces, where we find the same relation (\ref{cpxent}).
This relation (\ref{cpxent}) provides a new interpretation of areas of arbitrary convex surfaces on a time slice
in terms of the Euclidean path-integral complexity.\footnote{We would also like to note that there is another interpretation of $S(\Sigma)$ by a quantity called the differential entropy \cite{Dent}.}

\subsection{Holographic Path-Integral Complexity in Lorentzian AdS}\label{Lorentzian}

Now we would like to turn to the path-integral interpretation of Lorentzian AdS and its
path-integral complexity.
Here we encounter a new ingredient that the surfaces $M_\Sigma$ have variety of types:
space-like, null, and time-like. As we will see, studying the behavior of path-integral complexity
in the Lorentzian AdS will clarify the properties of their circuit interpretations.

As in the Euclidean AdS case, we can relate the normalization of wave functional for $|\Psi_\Sigma(M_\Sigma)\lb$
, defined by the path-integration on $M_\Sigma$, to the Lorentzian gravity action on $N_\Sigma$.
Here $N_\Sigma$ is the spacetime dual to the path-integration on $M_\Sigma$. When we consider a static gravity dual and choose $\Sigma$ to be on a canonical time slice $t=0$,
the spacetime $N_\Sigma$ is given by the region surrounded by $M_\Sigma$ and the slice $t=0$ (see Fig.\ref{fig:SSpathtime}).

The Euclidean path-integral complexity is defined by (\ref{Ecpxs}). In an analogous way, the Lorentizian path-integral complexity can be introduced as follow:
\ba
e^{iC(M_\Sigma)}=e^{iI_{G}(N_\Sigma)}. \label{phq}
\ea
Even though the phase factor of the wave functional initially has a phase shift ambiguity, we fixed this ambiguity by requiring that $C(M_\Sigma)=0$ for the trivial setup, which leads to the identification (\ref{phq}).

Also as in the Euclidean case, we may define the complexity of the state $|\Psi_\Sigma\lb$ by minimizing the gravity action $I_G(N_\Sigma)$:
\ba
C(|\Psi_\Sigma\lb)=\min_{M_\Sigma} C(M_\Sigma)=\min_{M_\Sigma}\left[I^L_{G}(N_\Sigma)\right].
\label{Lwww}
\ea

As a simple example, below we evaluate the path-integral complexity in the (Lorentzian version $t\to it$ of) the Poincare AdS background (\ref{pads}). Again we focus on $d=1$ i.e. AdS$_3$.
We set the surface $\Sigma$ to be the one at $z=\ep$ and $t=0$ extending in the $x$ direction.
We choose the surface $M_\Sigma$ to be a hyperplane which ends on $\Sigma$. We parameterize the time-like and space-like hyperplane, separately, as follows:
\ba
&& \mbox{Time-like hyperplane $M^{space}_\Sigma$}:\ t\sinh\theta+z\cosh\theta=0,  \label{msiga} \\
&& \mbox{Space-like hyperplane $M^{time}_\Sigma$}:\ t\cosh\ti{\theta}+z\sinh\ti{\theta}=0.
\label{msigb}
\ea
The limit $\theta\to\infty$ or $\ti{\theta}\to\infty$ makes the surface $M_\Sigma$ light-like.

When $M_\Sigma$ is space-like, we obtain:
\ba
I^L_G(M^{space}_\Sigma)&=&-\frac{1}{16\pi G_N}\int_{N_\Sigma} \s{g}(R-2\Lambda)
-\frac{1}{8\pi G_N}\int_{M_\Sigma} \s{h}K-\frac{1}{8\pi G_N}\int_{\Sigma}\s{\gamma}\left(\ti{\theta}+\frac{\pi}{2}i\right)\no
&=& \frac{cL}{12\pi\ep}\cdot\left[\frac{\sinh\ti{\theta}}{\cosh\ti{\theta}}-\ti{\theta}-\frac{\pi}{2}i\right].
\label{Lspace}
\ea
This result follows from the Euclidean result (\ref{ewww}) and (\ref{eww}) via the analytical continuation $\ti\theta=-i\ap$, $I^L_G=iI^E_G$ and $t^L=-it^E$. Interestingly this leads to the imaginary part of the corner contribution. This leads to
an exponentially large factor $e^{S(\Sigma)/4}$, identical to (\ref{cpxent}), in the total gravitational partition function $e^{iI^L_G}$. Thus we expect that the quantum circuit
on a space-like $M_\Sigma$ includes both Lorentzian (unitary) and Euclidean (non-unitary) gates.
Note that at $\ti{\theta}=0$, $M_\Sigma$ coincides with the canonical time slice. In this case the real part of $I^L_G(M^{space}_\Sigma)$ vanishes and its imaginary part agrees with the Euclidean AdS
result at $\ap=0$ (\ref{ewwa}). Since each of these two $M_\Sigma$ is an identical hyperbolic space,
we would like to argue the corresponding circuits are also the same, which includes only
 non-unitary gates.

On the other hand, when $M_\Sigma$ is time-like, the Lorentzian gravity action takes the form:
\ba
I^L_G(M^{time}_\Sigma)&=&-\frac{1}{16\pi G_N}\int_{N_\Sigma} \s{g}(R-2\Lambda)
+\frac{1}{8\pi G_N}\int_{M_\Sigma} \s{h}K-\frac{1}{8\pi G_N}\int_{\Sigma}\s{\gamma}\theta\no
&=& \frac{cL}{12\pi\ep}\cdot\left[\frac{\cosh\theta}{\sinh\theta}-\theta\right].\label{Ltime}
\ea
This is obtained from (\ref{Lspace}) via the analytical continuation $\theta=\ti{\theta}+\frac{\pi}{2}$. In this case the gravitational partition function becomes a
pure phase factor and thus we can conclude that the path-integration is Lorentzian (i.e. unitary).
When we take the limit $\theta\to 0$, where $M_\Sigma$ coincides with the AdS boundary $z=\ep$, the Lorentzian complexity gets equal to the Euclidean one (\ref{Ewwww}).

\begin{figure}
  \centering
  \includegraphics[width=7cm]{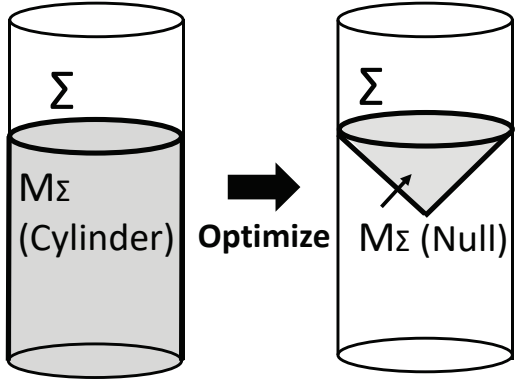}
 \caption{The optimization of Lorentzian path-integral into null one.}
\label{fig:optimizeC}
\end{figure}

If we adopt the definition of path-integral complexity for states dual to the Lorentzian AdS
(\ref{Lwww}), then the results (\ref{Lspace}) and (\ref{Ltime}) show that the minimum is realized in the null limit i.e. $\theta\to\infty$ or $\ti{\theta}\to\infty$. Interestingly, the fact that the complexity is minimized when the surface $M_\Sigma$ gets null seems to agree (up to a numerical factor $\frac{\pi}{2}$) with the ``complexity = action'' proposal \cite{Comp}, where the holographic complexity is given by the gravity action in the WDW patch.
Refer to Fig.\ref{fig:optimizeC} for this minimization. Indeed the WDW patch is identical to a double copy of $N_\Sigma$ in our setup. However note that in this limit $C(M_\Sigma)$ gets
negatively divergent for our gravity action. We expect that this difference comes from the treatment of null boundary and can be interpreted as the different choice of regularization of the null singularity.

 Also notice that it is not obvious if our path-integral complexity agrees with the ``complexity = action'' proposal for time-dependent gravity duals. Indeed, our path-integral complexity is computed only from the wave functional of $|\Psi_\Sigma\lb$ and its gravity dual $N_\Sigma$, which measures how much complicated preparing a given state is. On the other hand, the ``Complexity=Action'' proposal based on the WDW patch \cite{Comp} includes contributions of
gravity action from both before and after the quantum state is created.
Therefore, for example, in our approach, the time evolution of thermofield double state in two dimensional holographic CFTs, which is dual to the eternal BTZ  black hole \cite{MaE,HaMa}, seems to be computed from the gravity action on a spacetime which does not include the black hole singularity. The details will deserve future studies.\footnote{In actual computations in time-dependent backgrounds, we need to specify the boundaries of
the bulk region $N_\Sigma$. Clearly the past boundary should be the light-sheet as in the WDW patch prescription, which is true for our previous analysis for the Poincare AdS.
The choice of the future boundary is non-trivial and has to be specified in order to establish our new calculation. One natural choice will be the maximal time slice.
Another possibility is to take into backreactions which lead to a time-reversal symmetric gravity solution.}

\subsection{Path-Integral Circuit Complexity}\label{picc}

So far we focused on the holographic path-integral complexity for quantum states in CFTs.
It is also intriguing to consider a path-integral complexity for a unitary transformation itself.
Indeed, originally the computational complexity is defined for a unitary transformation as the minimum
number of gates which realize the unitary transformation.

\begin{figure}
  \centering
  \includegraphics[width=7cm]{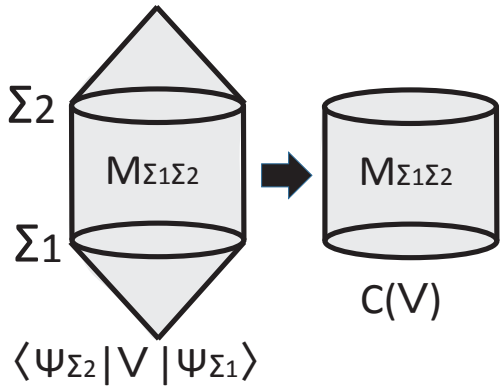}
 \caption{The gravity dual of path-integral complexity for the Hamiltonian evolution $C(e^{-iHt})$.}
\label{fig:complexityC}
\end{figure}

Consider a path-integral complexity
for the path-integral circuit defined by the codimension one surface $M_{\Sigma_1\Sigma_2}$
which connects the codimension two surfaces $\Sigma_1$ and $\Sigma_2$. We write this circuit
as $V[M_{\Sigma_1\Sigma_2}]$ and this evolves the state $|\Psi_{\Sigma_1}\lb$
into $|\Psi_{\Sigma_2}\lb$. In this setup, it is natural to identify
the path-integral complexity for this evolution $C(V[M_{\Sigma_1\Sigma_2}])$ as follows:
\ba
e^{C(|\Psi_{\Sigma_1}\lb)+C(|\Psi_{\Sigma_2}\lb)+C(V[M_{\Sigma_1\Sigma_2}])}=\la \Psi_{\Sigma_2}|V[M_{\Sigma_1\Sigma_2}]|\Psi_{\Sigma_1}\lb, \label{defcc}
\ea
where $C(|\Psi_{\Sigma_{1,2}}\lb)$ are the path-integral complexity of the states $|\Psi_{\Sigma_{1,2}}\lb$ $(\ref{Lwww})$ and the matrix element $\la \Psi_{\Sigma_2}|V[M_{\Sigma_1\Sigma_2}]|\Psi_{\Sigma_1}\lb$ is computed for the optimized states
$|\Psi_{\Sigma_{1,2}}\lb$ as in Fig.\ref{fig:complexityC}.

One of the simplest setups to calculate this circuit path-integral complexity will be for time-evolutions of holographic CFTs. In particular we choose the states $|\Psi_{\Sigma_{1,2}}\lb$
to be the CFT vacuum $|0\lb$ with the UV cut off scale $z=\ep$.
 The definition (\ref{defcc}) allows us to calculate its complexity $C(e^{-iTH})$ as the gravity action between $t=0$ and $t=T$. For the Poincare AdS$_3$ setup we can explicitly evaluate this as follows:
\ba
C(e^{-iTH})&=&-\frac{1}{16\pi G_N}\int_{N_{\Sigma_1\Sigma_2}} \s{-g}(R-2\Lambda)
-\frac{1}{8\pi G_N}\int_{M_{\Sigma_1\Sigma_2}}\s{-h}K \no
&=&-\frac{1}{4\pi G_N R^2_{AdS}}\int_{N_{\Sigma_1\Sigma_2}} \s{-g} +\frac{1}{8\pi G_N}
\int \s{-h}K  \no
&=& \frac{cTL}{12\pi\ep^2}.
\ea

We would like to note the obvious relation between our path-integral complexity for the Hamiltonian
evolution and the bare `energy' $E_{bare}(|\Psi\lb)$ of the state:
\be
\frac{d}{dt}C(e^{-itH})=E_{bare}(|\Psi\lb).  \label{llo}
\ee
This looks identical to the upper bound of the Lloyd bound \cite{LB} with a suitable redefinition of the complexity by a numerical factor.

Even though the above definition of path-integral complexity of a given unitary transformation is very natural from the field theory viewpoint,  we should notice that this quantity is not independent from the choice of quantum state on which the unitary operator acts. In the above example, we choose the vacuum state in a holographic CFT. In this sense, our quantity $C(V)$ seems to be different from the original definition complexity of the quantum circuit $V$.
One idea to extract such a universal part is to focus on the leading divergent term of $C(V)$, which is expected to be universal for any quantum states in the AdS/CFT. Another possibility to have a state independent holographic quantity for the circuit complexity is to employ the volume formula
\be
\sim\frac{1}{G_NR_{AdS}}\int_{M_\Sigma}\s{-g},  \label{volt}
\ee
which can be regarded as a time-like version of the ``complexity=volume'' conjecture in \cite{SUR,Susskind}, instead of the gravity action. It is obvious that this volume formula also leads to the same behavior
$C(e^{-itH})\sim \frac{cTL}{\ep^2}$.

We can also make the formula (\ref{volt}) slightly more covariant by replacing (\ref{volt}) with a modified  formula:
\ba
\sim\frac{1}{G_N}\int_{M_\Sigma\times I_{R_{AdS}}}\s{-g}{\cal L}_{G},
\ea
where ${\cal L}_{G}$ is the gravity action and $I_{R_{AdS}}$ is an interval
with the width $\sim R_{AdS}$, transverse to $M_\Sigma$.

As is clear from the above arguments, especially from the formula (\ref{volt}), the complexity of the Hamiltonian
circuit $C(e^{-itH})$ is proportional to $\s{-g_{tt}}$. In other words, the number of quantum gates in this circuit for a fixed time period $T$ determines the time component of the metric in the gravity dual. More explicitly, if we choose $M_\Sigma$ to be a surface at $z=z_0$ and take the range of $x$ to be the UV cut off scale $L=z_0$, then we find
\be
C(V_{\Sigma})\sim \frac{cT}{z_0},   \label{cvcut}
\ee
which agrees with $\int^T_0 dt\s{g_{tt}}$ at $z=z_0$.
In this way, this result shows that the non-zero metric component $g_{tt}$ emerges from the non-zero density of unitary quantum gates in the Hamiltonian circuit. If we consider a trivial quantum system with the Hamiltonian proportional to the identity $H\propto 1$, then we expect from the above arguments that in its gravity dual, we have $g_{tt}=0$, i.e. the time coordinate is vanishing. In this way, we have reached the idea that the time coordinate in a gravity dual emerges from the complexity of the Hamiltonian circuit in the dual field theory. Our later arguments in the next section further support this idea.

\section{Entanglement Evolutions in Path-Integral Circuits} \label{evo}

Now we would like to turn to dynamics of quantum entanglement for the path-integral circuits.
Again we will study the Euclidean and Lorentzian AdS setups separately below.

\subsection{Entanglement Evolutions in Euclidean AdS}

\begin{figure}
  \centering
  \includegraphics[width=7cm]{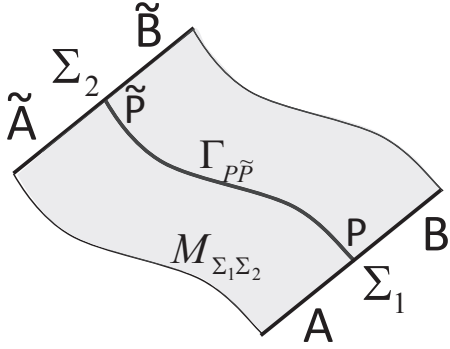}
 \caption{The path-integration on $M_{\Sigma_1\Sigma_2}$ which connects $\Sigma_1$ to $\Sigma_2$, which is regarded as a quantum circuit.}
\label{fig:EEpath}
\end{figure}

Consider a codimension one surface $M_{\Sigma_1\Sigma_2}$ in an Euclidean AdS, which connects two codimension two surfaces $\Sigma_1$ and $\Sigma_2$. We divide $\Sigma_1$ and $\Sigma_2$ into $A,B$ and $\ti{A},\ti{B}$, respectively as in Fig.\ref{fig:EEpath}. Our conjecture argues that its corresponds to a (non-unitary) quantum circuit  which maps $|\Psi_{\Sigma_1}\lb$ into $|\Psi_{\Sigma_2}\lb$. For explicit construction of such circuits in the context of AdS$_3/$CFT$_2$ refer to  the appendix B.

The codimension three surfaces which separate $A,B$ and $\ti{A},\ti{B}$ are called $P$ and $\ti{P}$, respectively. $\Gamma_{P\ti{P}}$
is a codimension two surface which connects $P$ and $\ti{P}$. We are mainly focusing on the local geometry around $\Gamma_{P\ti{P}}$.

We are interested in how quantum entanglement is produced by the circuit evolution along  $\Gamma_{P\ti{P}}$. To quantify this we consider the entanglement entropy $S_{A\ti{A}}$ defined by
\ba
S_{A\ti{A}}=-\mbox{Tr}[\rho_{A\ti{A}}\log\rho_{A\ti{A}}],
\ea
where $\rho_{A\ti{A}}$ is the reduced density matrix
\be
\rho_{A\ti{A}}=\mbox{Tr}_{B\ti{B}}[|\Psi_{\Sigma_1\Sigma_2}\lb\la\Psi_{\Sigma_1\Sigma_2}|].
\ee
The pure state $|\Psi_{\Sigma_1\Sigma_2}\lb$ is obtained by the path-integrations on $M_{\Sigma_1\Sigma_2}$ via the channel-state duality (see e.g.\cite{Hosur:2015ylk,CJQW}).
Equivalently, we write $V[M_{\Sigma_1\Sigma_2}]$ as the (non-unitary) quantum circuit corresponds to the Euclidean path-integration on $M_{\Sigma_1\Sigma_2}$. Then the pure state $|\Psi_{\Sigma_1\Sigma_2}\lb$ is defined by
\ba
|\Psi_{\Sigma_1\Sigma_2}\lb={\cal{N}}\cdot \sum_{i}\left(V[M_{\Sigma_1\Sigma_2}]|i\lb_{\Sigma_1}\right)\otimes |i\lb_{\Sigma_1}, \label{evole}
\ea
where $|i\lb_{\Sigma_{1,2}}$ is the complete basis of the Hilbert space for $\Sigma_{1,2}$ and ${\cal N}$ is the normalization constant. It is convenient to choose the real space basis for $|i\lb$.
Since we consider space-like path-integrations, $V[M_{\Sigma_1\Sigma_2}]$ becomes non-isometric.
We also would like to mention that this entanglement entropy $S_{A\ti{A}}$ for a given circuit $V$ is essentially
the same as the quantity called operator entanglement entropy studied in \cite{Dubail,Zhou,NoRy}.

In particular, when $M_{\Sigma_1\Sigma_2}$ is squeezed to zero size, we simply find
$S_{A\ti{A}}=0$ because $V[M_{\Sigma_1\Sigma_2}]=I$. Thus if we perform any generic path-integrations  $V[M_{\Sigma_1\Sigma_2}]\neq I$, then we expect $S_{A\ti{A}}$ increases at least initially. Motivated by this we would like to focus on the case where the evolving surface $M_{\Sigma_1\Sigma_2}$ is infinitesimally short, i.e. $\Sigma_1$ and $\Sigma_2$ are very closed to each other.
In this case we can regard
$\Gamma_{P\ti{P}}$ as an (infinitesimally short) extremal surface which connects $P$ and $\ti{P}$.
Indeed, in this case we can ignore the global geometry and focus on the local geometry near $\Gamma_{P\ti{P}}$. In this setup we conjecture\footnote{Here in the formula (\ref{basic}) we implicity
ignore contributions from non scrambling quantum gates such as the dilation. We leave this issue until section 4.3. as it gets much clear in Lorentzian setups.} the following relation between the infinitesimal growth of entanglement entropy, denoted by $dS_{A\ti{A}}$, and the infinitesimal area of $\Gamma_{P\ti{P}}$, denoted by $dA(\Gamma_{P\ti{P}})$:
\be
dS_{A\ti{A}}=\frac{dA(\Gamma_{P\ti{P}})}{4G_N}.  \label{basic}
\ee
This formula offers an interpretation of an arbitrary area element in the Euclidean AdS
in terms of quantum entanglement evolutions.

Another important property is that $S_{A\ti{A}}$ depends only on $\Sigma_1$ and $\Sigma_2$, while
it is independent from the choice of $M_{\Sigma_1\Sigma_2}$. This is because we can equivalently
deform the integration manifold by the Weyl rescaling as in our previous arguments.

\begin{figure}
  \centering
  \includegraphics[width=8cm]{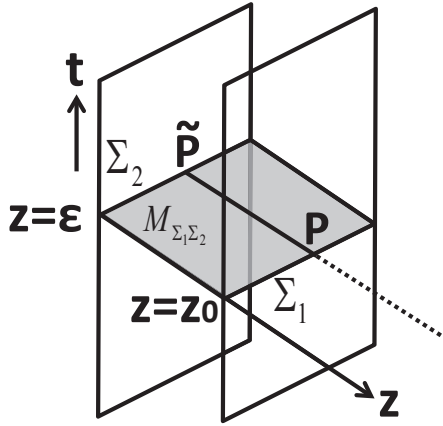}
 \caption{An examples of space-like path-integration in a Poincare AdS$_3$.}
\label{fig:EEpatth}
\end{figure}

As a simple example of space-like path-integrations, consider the case where $\Gamma_{P\ti{P}}$ is a straight line geodesic in Poincare AdS$_3$, depicted in Fig.\ref{fig:EEpatth}. We choose
$\Sigma_2$ at the AdS boundary $z=\ep$ such that $|\Psi_{\Sigma_2}\lb$ is the CFT vacuum and $\Sigma_1$ is at $z=z_0$. Then the length of $\Gamma_{P\ti{P}}$ is computed as
\be
A(\Gamma_{P\ti{P}})=R_{AdS}\int^{z_0}_\ep\frac{dz}{z}=R_{AdS}\log\frac{z_0}{\ep}.
\label{ageo}
\ee
Thus we get
\be
dS_{A\ti{A}}=\frac{c}{6}\cdot \frac{dz}{z},  \label{difee}
\ee
where $c$ is the central charge of the 2d CFT. If we set $z_0$ to infinity or some IR cut off length
$\xi$, $S_{A\ti{A}}$ clearly agrees with the standard result of entanglement entropy in 2d CFT \cite{CaCa}:
$S_A=\frac{c}{6}\log \frac{\xi}{\ep}$. This is because when $z_0$ gets infinite, the surface $\Sigma_1$ shrinks to zero size and  the IR subsystem $A$ disappears. The infinitesimally contribution (\ref{difee}) is also naturally interpreted as the entanglement production due to the quantum gates which intersect with the line segment $\Gamma_{P\ti{P}}$. In this special setup we expect the integrated form is also correct:
\ba
S_{A\ti{A}}=\frac{A(\Gamma_{P\ti{P}})}{4G_N}=\frac{c}{6}\log \frac{z_0}{\ep}.  \label{integee}
\ea
Later in subsection \ref{ghost} we will present another interpretation of $S_{A\ti{A}}$ as a holographic entanglement entropy.

However in generic setups, the integrated formula (\ref{integee}) is not correct.
Since the circuit $M_{\Sigma_1\Sigma_2}$ is not optimal in general when the distance
between $\Sigma_1$ and $\Sigma_2$ is finite, we expect
\ba
S_{A\ti{A}}\leq \frac{A(\Gamma_{P\ti{P}})}{4G_N},
\ea
for generic choices of $M_{\Sigma_1\Sigma_2}$.
For example, a typical such example will be a path-integration over a Euclidean time interval $0\leq t\leq T$ at the AdS boundary $z=\ep$. The straight time evolution on $M_{\Sigma_1\Sigma_2}$ leads to
\be
\frac{A(\Gamma_{P\ti{P}})}{4G_N}=\frac{c T}{6\ep}. \label{saap}
\ee
However, since this state is given by the thermofield double state (or equally $V[M_{\Sigma_1\Sigma_2}]=e^{-TH}$):
\be
|\Psi_{\Sigma_1\Sigma_2}\lb={\cal N} \sum_i e^{-TH}|i\lb_{\Sigma_1}|i\lb_{\Sigma_2},  \label{tfds}
\ee
where the CFT Hamiltonian $H$ acts only on $|i\lb_{\Sigma_1}$.
 The gravity dual of this state is given by a lower half of BTZ black hole \cite{MaE,HaMa}, which leads to
the etimation
\be
S^{BTZ}_{AA}\simeq \frac{c}{3}\log\frac{2T}{\pi\ep}. \label{btzgeo}
\ee
Therefore we clearly find\footnote{One might be tempting to argue
that the minimum of $\frac{A(\Gamma_{P\ti{P}})}{4G_N}$ over all possible choices of $\Gamma_{P\ti{P}}$
can be equal to $S_{A\ti{A}}$. However, this is not exactly correct in general, though this seems to give a good approximation. Indeed, in the present example given by the state (\ref{tfds}), the minimum of the geodesic length in the pure AdS is given by $\frac{c}{3}\log\frac{T}{\ep}$ which is larger than the
BTZ result (\ref{btzgeo}) by $\log \frac{\pi}{2}>0$.}
$S_{A\ti{A}}<\frac{A(\Gamma_{P\ti{P}})}{4G_N}$. On the other hand, when $T$ is infinitesimally small\footnote{One may worry that if $T\ll \ep$, then the interval $\Gamma_{P\ti{P}}$ does not include
any quantum gate. However, the actual lattice spacing in a holographic CFT with a large central charge $c\gg 1$, is expected to be $\ep/c$ taking into account a `fractionalization', which will be discussed in the section 5. Therefore we can consider the parameter region $\ep/c\ll T\ll \ep$ to
have a sensible result.} $T\ll \ep$, we can trust the estimation (\ref{saap}) and this agrees with our conjecture (\ref{basic}).

Finally, it is also intriguing to consider the another quantity $S_{\ti{A}}-S_{A}$, which simply measures
the growth of the entanglement entropy by comparing the initial state $|\Psi_{\Sigma_1}\lb$ with
the final state $|\Psi_{\Sigma_2}\lb$. However, in this case
there is already non-zero entanglement i.e. $S_A>0$ for the initial state $|\Psi_{\Sigma_1}\lb$ and it is not clear how efficiently the quantum gates along $\Gamma_{P\ti{P}}$ add quantum entanglement. Therefore we expect that the surface area of $\Gamma_{P\ti{P}}$ gives an upper bound
\ba
S_{\ti{A}}-S_A\leq \frac{A(\Gamma_{P\ti{P}})}{4G_N}.
\ea
For example, if we consider the case of (\ref{ageo}) in Fig.\ref{fig:EEpatth}, this inequality is saturated, while it is not when $\Gamma_{P\ti{P}}$ is tilted.

\subsection{Ghost D-brane Holography and Calculation of  $S_{A\ti{A}}$}\label{ghost}

Before we go on, let us drop by another interpretation of the quantity $S_{A\ti{A}}$ .
 We would like to present a new holographic setup where we can directly calculate $S_{A\ti{A}}$ introduced
 as in Fig.\ref{fig:EEpatth} in the previous subsection.\footnote{Since this subsection is independent from main contents of this paper, readers can skip this part at first.} For this we would like to identify the dual CFT description of a part of Poincare AdS defined by $\ep\leq z \leq z_0$. We argue this is dual to a supergroup $SU(N|N)$ gauge theory. Its ghost sector has the degrees of freedom with a length scale shorter than $z_0$, while the regular gauge theory part is defined up to the UV cut off scale $\ep$.

If we consider the AdS$_5\times$ S$^5$ as a concrete setup of AdS/CFT in string theory, such a supergroup gauge theory appears if we consider $N$ D3-branes and $N$ ghost D3-branes \cite{OT}. A ghost D-brane is an object which simply annihilates a D-brane without leaving any radiations or backreactions (thus, is different from an anti D-brane). In the language of the boundary states in boundary conformal field theories, the boundary state for a ghost D-brane is just given by $-|B\lb$, if we write
 that for a standard D-brane as $|B\lb$. Since a D3 and a ghost D3 can simply annihilate, the partition function of the supergroup $U(N|M)$ gauge theory is equal to that of
$U(N-M)$ gauge theory. Therefore in the present setup we can completely annihilate degrees freedom for the IR length scale $z>z_0$. See \cite{EMR} for a similar but different way to use the supergroup gauge theory as the change of cut off scale.

Here we are interested in the computation of entanglement entropy when we divide the system into two parts $A$ and $B$ in this supergroup gauge theory. We expect that the real gauge theory degrees of freedom live on the original AdS boundary $z=\ep$, while the ghost degrees of freedom on
the new boundary $z=z_0$. Therefore it is clear that the holographic entanglement entropy precisely agrees with $S_{A\ti{A}}$ in (\ref{integee}).

It is also intriguing to note that such a holography with ghost D-branes has a lot of applications.
For example, consider again an AdS/CFT setup of the AdS$_5\times$ S$^5$ dual to the gauge theory on $N$ D3-branes.
We take the Euclidean Poincare metric (\ref{pads}) for AdS$_5$. Let us couple the ghost gauge fields, which come from $N$ ghost D3-branes, localized on a disk $t^2+|x|^2\leq l^2$ in the dual CFT$_4$.
  Then its holographic dual can be identified with the Euclidean Poincare AdS$_5$ with a half ball, defined by $t^2+|x|^2+z^2\leq l^2$, removed. This spacetime can be regarded as the original AdS$_5$ minus the holographic dual of the BCFT \cite{AdSBCFT} dual to the ghost fields.  In this way, a local coupling of ghost D-branes can eliminate some part of the holographic spacetime in general.

As a final example, consider the AdS$_3/$CFT$_2$ with a Poincare AdS and introduce ghost fields localized on an interval $0\leq x \leq l$ at any time $t$. Clearly, the dual CFT$_2$ lives on two disconnected
half lines $x\leq 0$ and $x\geq l$. In the gravity dual, we expect these two
disconnected boundaries are connected in the bulk, as the total geometry should be given by a Poincare AdS$_3$ with a half solid cylinder removed. Thus this geometry provides a new example of traversal wormholes, different
from the construction in \cite{GJW}. Indeed the ghost degrees of freedom in CFT$_2$ can violate the null energy condition in the gravity dual.

It will be a very intriguing future problem to study the new holography in more details.

\subsection{Entanglement Evolutions in Lorentzian AdS}

Now we move on to the dynamics of quantum entanglement in Lorentzian AdS setups.
Consider the evolution of states as in Fig.\ref{fig:EEpath}. Again things are more complicated than the Euclidean ones because the surface $M_\Sigma$ can be
either space-like, null or time-like.

One important hint to understand how their circuits look like is our previous calculation of path-integral complexity in section \ref{PathC}. As we have found there, when $M_\Sigma$ is space-like, the corresponding circuits consist of both unitary gates and non-unitary gates.\footnote{Note that the presence of unitary gates is important as supported from our previous result of path-integral complexity (\ref{Lspace}). Indeed, if there were only non-unitary gates, then it would suggest that there are more optimized Euclidean circuit than the one dual to the canonical time slice
in the static case. This contradicts with the optimization of circuits for Euclidean AdS.} On the other hand, when $M_\Sigma$ is time-like, the circuits consist only of
unitary Lorentzian gate.

However, one might still be puzzled by the fact that the area of
$\Gamma_{P\ti{P}}$ can be vanishing when $M_\Sigma$ is null
if we naively extend the formula (\ref{basic}) to the current Lorentzian setup.
 This is because we can easily find
an example where the entanglement entropy is growing $S_A-S_{\ti{A}}>0$ even though the surface
$\Gamma_{P\ti{P}}$ is null. For example, we can consider the setup of Fig.\ref{fig:SSpathtime} for
Lorentzian Poincare AdS and choose the surface $M_\Sigma$ to be null i.e. $t+z=0$.
This argument shows that even though $\Gamma_{P\ti{P}}$ has the zero area, it creates non-vanishing entanglement. At first sight this looks contradicting with our interpretation that the area is related to the number of quantum gates.

\begin{figure}
  \centering
  \includegraphics[width=8cm]{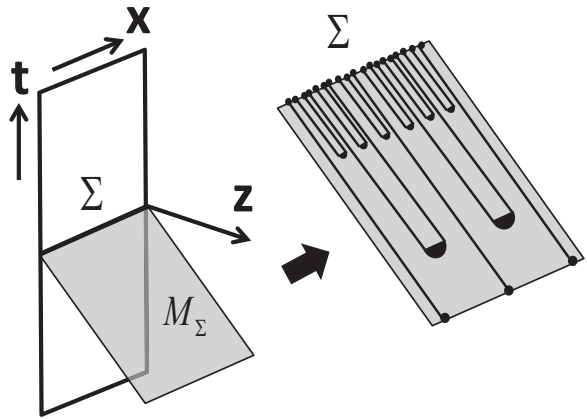}
 \caption{A sketch of a quantum circuit dual to a null surface. Strictly speaking we need to
 maintain the translational invariance with suitable rearrangements.}
\label{fig:NullC}
\end{figure}

We would like to argue that this paradox can be resolved if we think null circuits consist of not only
the trivial gates (i.e. the identity $I$ transformation) but also the pair creations each of which create extra dimension of Hilbert space as sketched in Fig.\ref{fig:NullC}. An important point is that such quantum gates does not scramble the quantum system. Indeed, from the causal structure of the AdS, we expect the excitations do not spread in a relativistically but simply propagates vertically along the null-lines. We claim that the area element does not include contributions from such non-scrambling gates.

In the appendix B, we analyze explicit examples of path-integral circuits, which correspond to the case where $M_\Sigma$ is given by de-Sitter space or hyperbolic space in AdS$_3$, applying the results in the recent work \cite{Path}. In this analysis, the non-scrambling quantum gates correspond to the dilation $L'$ and the scrambling ones to the Hamiltonian $H_0$ of the CFT. The dilation does not scramble excitations as can be seen from the operator transformation (\ref{localt}).

Motivated by these observations, we would like to propose the following relation for infinitesimally small $\Gamma_{P\ti{P}}$:
\ba
\left(dS^{s}_{A\ti{A}}\right)^2-\left(dS^{t}_{A\ti{A}}\right)^2
=\left(\frac{dA(\Gamma_{P\ti{P}})}{4G_N}\right)^2,   \label{eee}
\ea
where our definition of each area is always given by $A(\Gamma_{P\ti{P}})=\int_{\Gamma_{P\ti{P}}} \s{g}$ such that the area becomes imaginary for time-like surfaces. $dS^{s}_{A\ti{A}}$ and $dS^{t}_{A\ti{A}}$ each describes the increased amount of entanglement entropy due to the scrambling non-unitary (Euclidean) and unitary (Lorentzian) gates along $\Gamma_{P\ti{P}}$, respectively. As we mentioned just before, we do not take into account the contributions from non-scrambling quantum gates like pair creations. One possibility to make our definition of $dS^{s,t}_{A\ti{A}}$ more explicit may be to use the quantity called tripartite mutual information
\be
I_3(A:B:\ti{A})=S_A+S_B+S_{\ti{A}}-S_{AB}-S_{A\ti{A}}-S_{\ti{A}B}+S_{AB\ti{A}},
\ee
which measures the amount of scrambling \cite{Hosur:2015ylk}. This quantity is known to be non-positive in holographic entanglement entropy \cite{Hayden:2011ag}.
 Therefore we may define $dS^{s,t}_{A\ti{A}}$ to be the growth of $-I_3(A:B:\ti{A})$ due to non-unitary or unitary gates.

\begin{figure}
  \centering
  \includegraphics[width=7cm]{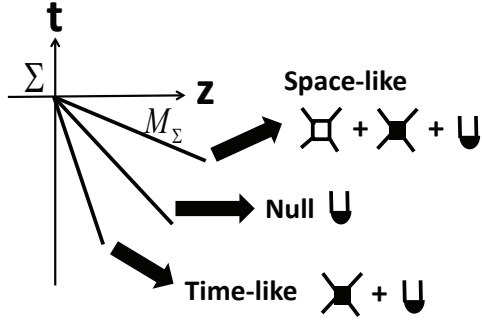}
 \caption{The structures of quantum gates for each slice $M_\Sigma$ in the Poincare AdS, where $M_\Sigma$ is either space-like, null or time-like. The white square gate with four legs corresponds to a non-unitary (Euclidean) one, while the black one to a unitary (Lorentzian) one. The black semi circle gate with two legs describes a pair creation of EPR pair. The first two quantum gates contribute to
 the left hand side of (\ref{eee}), while the final one does not.}
\label{fig:complexityLQC}
\end{figure}

If there are no unitary circuits as is true for the canonical time slice in a static gravity dual, the above formula is reduced to the previous one (\ref{basic}) for Euclidean setups with the identification $dS_{A\ti{A}}=dS^{s}_{A\ti{A}}$. On the other hand, for time-like surfaces $M_\Sigma$, the space-like part is vanishing $dS^{s}_{A\ti{A}}=0$. Refer to Fig.\ref{fig:complexityLQC} for a sketch of this interpretation.

In time-dependent backgrounds, the holographic entanglement entropy is given by
the extremal surface area \cite{HRT}. Therefore we expect $dS^{t}_{A\ti{A}}=0$ when $\Gamma_{P\ti{P}}$ is a part of a extremal surface. This observation enables us to decompose the area into
$dS^{t}_{A\ti{A}}$ and $dS^{s}_{A\ti{A}}$.\\

\begin{figure}
  \centering
  \includegraphics[width=7cm]{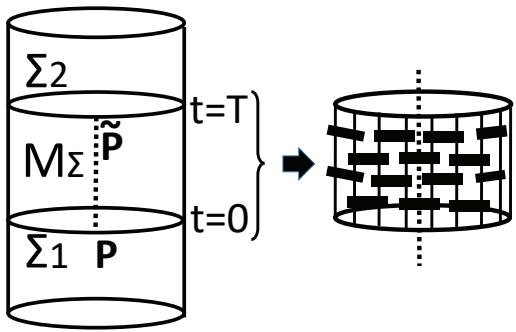}
 \caption{An example of entanglement evolution in a time-like path-integral in a Lorentzian setup.
 In the left picture describes its gravity analysis. The right picture is a sketch of its dual quantum circuit,
 where the vertical lines represent qubits and the thick horizontal intervals denote the unitary quantum gates.
 We count the number of gates which intersect with the dotted vertical line.}
\label{fig:timeEV}
\end{figure}

As a simple example of the unitary circuits, consider the real time evolution
of a CFT dual to the Poincare AdS$_{d+2}$ as depicted in Fig.\ref{fig:timeEV}.
The surface $\Sigma$ is chosen to be $R^d$:
$-\infty<x_1,\ddd,x_d<\infty$ at $z=\ep$ and $t=0$. We choose $P$ and $\ti{P}$ are separated by $T$ in the $t$ direction and are both at $x_1=0$. The relation (\ref{eee}) leads to
\be
dS^{t}_{A\ti{A}}=\frac{R^d}{4G_N}\cdot \frac{L^{d-1}}{\ep^d}\cdot dT.
\ee
For 2d CFTs, in particular, we find
\be
dS^{t}_{A\ti{A}}=\frac{c}{6}\cdot \frac{dT}{\ep}.  \label{tdeeg}
\ee
Obviously in this case there is no non-unitary contributions $dS^{s}_{A\ti{A}}=0$.

It is intriguing to notice that the estimation (\ref{tdeeg}) of the number of unitary gates agrees with our
previous one (\ref{cvcut}) obtained from the holographic path-integral complexity. This supports the idea that
the time component of the metric in gravity duals emerges from the density of (scrambling) unitary quantum gates as we argued in section \ref{picc}.

We would like to compare this prediction with the CFT calculation. The entanglement entropy produced by the
unitary evolution $V=e^{-iTH}$ can be measured as in (\ref{evole}) by considering the evolution:
\be
|\Psi_{\Sigma_1\Sigma_2}(T)\lb={\cal N}\cdot e^{-iTH}e^{-\beta H/2}\sum_{i}|i\lb_{\Sigma_1}|i\lb_{\Sigma_2}.
\ee
Here we regularize by inserting a damping factor $e^{-\beta H/2}$, where we cut off the length scale shorter than $\beta(\ll 1)$ and thus we expect $\beta\sim \ep$. Then this is the same as the evolution of entanglement entropy of thermofield double state in 2d CFTs \cite{ANT,HaMa}:
\be
S_A(T)=\frac{c}{3}\log\left[\frac{\beta}{\pi\ep}\cosh\frac{\pi T}{\beta}\right]\simeq \frac{\pi cT}{3\beta},
\ee
where we assumed $T\gg \beta$ because $\beta$ is a regularization parameter. Indeed, this agrees with
(\ref{tdeeg}) up to an undetermined $O(1)$ constant factor. Thus, this supports our conjectured relation (\ref{eee}).

Let us emphasize that the relation (\ref{eee}) offers a quite new calculation of entanglement entropy in AdS/CFT because it relates the area of time-like surface to a growth of entanglement entropy, though the relation for a purely space-like surface can be regarded as a large generalization of the idea of holographic entanglement entropy \cite{RT,HRT}.

\subsection{Gravitational Force from Quantum Circuits}

As our final argument, we would like to consider how gravitational force can be understood in terms of
the quantum circuits. For this, imagine a point particle with a mass $m$ in a three dimensional gravity dual. Since it propagates along a time-like trajectory, which is chosen to be $\Gamma_{P\ti{P}}$,
our formula (\ref{eee}) argues
\be
dS^{t}_{A\ti{A}}=\frac{1}{4G_N}\int^{\ti{P}}_P \s{-g}.  \label{qqqq}
\ee
Note that the left hand side counts the number of scrambling quantum gates which act when the particle moves on the interval $\Gamma_{P\ti{P}}$. If we consider a static particle $x=$const. in
a weak gravitational potential $\phi(x)\ll 1$ and focus its neighborhood approximated by a flat space, we have
$g_{tt}\simeq -(1+2\phi(x))$. Therefore we obtain the estimation
\ba
[\#\ \mbox{of scrambling quantum gates on}\ \Gamma_{P\ti{P}}] \simeq \frac{1}{4G_N}\int^{\ti{P}}_P dt(1+\phi(x)).
\label{gatepot}
\ea

Now we would like to remember that a point particle is a localized excitation in a quantum circuit.
For more explanations, refer to \cite{Numa}, where such a connection has been discussed in the context of tensor network and holography.\footnote{As argued in the paper \cite{Numa}, from the viewpoint of quantum entanglement, we can understand the gravitational force in AdS as a sort of decoherence as follows. A localized excitation in a CFT can spread out at the speed of light. If we consider a tensor network description (like MERA) of this excited state in the CFT, initially the locally excited state is expressed as a tensor network where the tensor in the UV layer is locally modified from that for the CFT vacuum. As the time evolves, the location of such a modified tensor moves toward the internal layer
such that the length scale $z$ of the modified layer is approximated by the time $t$. This explains the light-like trajectory $z\simeq t$ in the dual Poincare AdS.} Thus the presence of scrambling quantum gates gives an obstruction to preserve the localized form of such an excitation.
If it experiences a lot of scrambling quantum gates, it can be spread over in a wide region as the gravitational wave radiations. This motivates us to argue that the preferred direction in which the particle tends to propagate, is the one with smaller number of quantum gates. In order words, the particle moves so that it decreases the value of the gravitational potential $\phi(x)$ in (\ref{gatepot}), as expected from the standard mechanics. This also explains why accelerated particles eventually
approach to null geodesics as there are no active gates along the null surfaces (refer to Fig.\ref{fig:NullC}).

We can go further and make a speculative argument that by counting the number scrambling quantum gates we can
understand the standard particle action itself
\be
I_p=-m\int dt\s{-g}.
\ee
Indeed, we can understand the phase factor $e^{iI_p}$ if we assume that for each gate we get the same
phase factor $e^{i\theta}$, because the density of unitary gate along a time-like trajectory is proportional to $\s{-g}$ as follows from our formula (\ref{qqqq}). The wave function of the particle located at
$z=z_0$ in the Poincare AdS$_3$ behaves as
\be
\psi\sim e^{-im\int^P_{\ti{P}} dt\s{-g}}\simeq e^{-i\Delta\frac{T}{z_0}}, \label{phasew}
\ee
where $\Delta$ is the conformal dimension and $z_0$ is the effective lattice spacing and $T$
is the time period between $P$ and $\ti{P}$. Indeed, the number of quantum gates between $P$ and $\ti{P}$ is estimated as $\frac{T}{z_0}$ and the above behavior (\ref{phasew})
in the gravity dual is explained if we set $e^{i\theta}=e^{-i\Delta}$, which is quite natural
in the light of the Lloyd bound (\ref{llo}). In this way, our interpretation of the gravity dual as a collection of quantum circuits enables us to explain the particle action. As usual, the semiclassical approximation of the path-integrations over particle trajectories $\int Dx \ e^{iI_p}$ leads to the equation of motion $\delta I_p=0$, i.e. the motion of a massive particle with a gravitational force.
Notice that in the CFT side, the wave function $\psi$ represents only a localized excitation part of the whole wave function of the quantum many-body system, given by a discretization of the CFT.

We can generalize these arguments
to higher dimensions by dividing the area of $\Gamma_{P\ti{P}}$ in (\ref{eee}) by an UV cut off (lattice spacing).

\section{Conclusions and Discussions}

In this article, we presented a proposal that a gravity dual spacetime consists of quantum circuits such that each surface $M_\Sigma$ in the spacetime corresponds to a quantum circuit defined by a
 path-integration on $M_\Sigma$ with a suitable UV cut off. Our construction was achieved by
 developing the surface/state correspondence \cite{MiTa}. Our proposal covariantly generalizes and refines the earlier conjectures which relate emergent spaces in AdS/CFT to various tensor networks, which have been restricted to specific slices such as canonical time slices. We believe that our proposal gives a simple summary of what we expect for the connection between the AdS/CFT and tensor networks and that it is one of key principles of holography.
 Our arguments can be applied to both Euclidean and Lorentzian asymptotically AdS backgrounds.
A table which briefly summarizes our holographic relations is as follows:
\ba
&& \bullet\  \mbox{The bulk region $N_\Sigma$ surrounded
 by a codimension one surface $M_\Sigma$}  \no
&&\ \ \ \ \ \ \ \ \ \ \ \ \ \ \ \ \ \lr \mbox{The quantum circuit $V_\Sigma$ defined by a path-integral on $M_\Sigma$} \no
&& \bullet\ \mbox{(The bulk region surrounded
 by) a codimension two surface $\Sigma$} \no
&&\ \ \ \ \ \ \ \ \ \ \ \ \ \ \ \ \ \lr
\mbox{The quantum state $|\Psi_\Sigma \lb$}\ \ \   \mbox{(i.e. the surface/state correspondence)} \no
&&\bullet\ \mbox{The gravitational action on $N_\Sigma$} = \mbox{The path-integral complexity of the circuit $V_\Sigma$} \no
&&\bullet\ \mbox{The area of codimension two infinitesimally small surface $\Gamma_{P\ti{P}}$}\no
&& \ \ \ \ \ \ \ = \mbox{The number of scrambling quantum gates which intersect with $\Gamma_{P\ti{P}}$.} \nonumber
\ea

We studied several outcomes of our proposal from the viewpoint of complexity and entanglement entropy.
We argued that a holographic counterpart of the path-integral complexity can be computed from the gravity action restricted to suitable regions.

In Euclidean gravity duals, we found that the minimum of holographic path-integral complexity, identified with the complexity of a quantum state, is dominated by the corner contribution which is equal to the surface area. The qualitative behavior of our Euclidean holographic complexity looks similar to the ``complexity = volume'' proposal \cite{SUR,Susskind}, though not exactly the same.
Moreover, this provides a new interpretation of a generic surface area in gravity duals.

In Lorentzian gravity duals, we evaluated the holographic path-integral complexity and found that this reproduces
the holographic complexity of ``complexity = action'' \cite{Comp}, given by the gravity action in a WDW patch if the background is static. This provides the first derivation of holographic complexity from our basic principle of holography. For time-dependent backgrounds, our holographic results of path-integral complexity seem to deviate from the earlier proposal of holographic complexity, which will require a future analysis. This computation of path-integral complexity also clarified the structures of quantum gates for each surfaces $M_\Sigma$ in a Lorentzian AdS. We also defined and evaluated the path-integral complexity for unitary operators.

The analysis of quantum entanglement in our framework reveals a direct connection between the number of scrambling gates and the surface area. This relation is simple for Euclidean setups and can be regarded as a natural generalization of holographic entanglement entropy \cite{RT,HRT}. We also pointed out that this new quantity can sometimes be regarded as
a holographic entanglement entropy in a ghost D-brane holography. However, for Lorentzian gravity duals, the connection gets more non-trivial due to the presence of both unitary and non-unitary quantum gates, summarized by the formula (\ref{eee}).

It is also intriguing that our results for the complexity and quantum entanglement show that the time component of the metric in AdS emerges from the density of scrambling unitary quantum gates in the dual CFT. This largely reinforces the idea of emergent space from quantum entanglement so that it includes the time coordinate. We also gave a heuristic argument how the gravitational force is explained from the viewpoint of our quantum circuit picture.

Also we would like to comment on the UV cut off or lattice spacing in our formulation.
In holographic CFTs with classical gravity duals \cite{HKS}, we expect an extra property of UV cut off such that
the bulk gravity becomes local in a length-scale much shorter than the AdS radius. For AdS$_3/$CFT$_2$,
we expect that the actual lattice spacing in the 2d CFT is fractionalized to be $\ep/c$, where $\ep$
is the original one.  Consider a symmetric product CFT on a circle with the radius $R_0$ defined by $n$ copies of a seed CFT as a typical example of CFTs with holographic duals. Its long string
sector, which dominates the degrees of freedom, behaves like a CFT on a larger cylinder with the radius
$nR_0$ \cite{MaSuF}. Therefore it has a fractionalized momentum which matches with the above mentioned fined grained
lattice spacing \cite{MTW}. In $d+1$ dimensional CFTs, we similarly expect that the actual lattice spacing
looks like $ \ep/c^{1/d}$, where $c$ denotes the central charge defined by $R^d_{AdS}/G_N\sim c$.
For a $(2+1)$ dimensional $U(N)$ gauge theory on a torus $T^2$, we can define the long string sector by
the twisted boundary condition $\Phi(x+2\pi R_0)=U\Phi(x)U^{-1}$ and $\Phi(y+2\pi R_0)=V\Phi(y)V^{-1}$ with the $N\times N$ matrices
$U$ and $V$  such that $UV=VUe^{\frac{2\pi i}{N}}$. This again leads to the fractionalizations of the momenta
by $1/N$, indeed leading to the advertised lattice spacing $\ep/N\sim \ep/\s{c}$.

There are many problems we would like to explore in future works.
We would like to explore constructions of the quantum circuits from our path-integrations and
consider their connections to existing tensor networks. We also need to understand how the dynamics of
Einstein equation directly emerges from our picture. It is also intriguing to study more details of our
holographic path-integral complexity including time-dependent gravity duals. Finally, it is very important
to investigate implications of our formulation in non-AdS spacetimes such as de-Sitter spaces
(refer to \cite{DST} for a recent interesting argument and see also  \cite{Narayan,Sato,MTW}
for earlier related discussions).

\subsection*{Acknowledgements}
 We thank Arpan Bhattacharyya, Michael Heller, Alexander Jahn, Romould Janik, Yuya Kusuki, Tomoyuki Morimae, Robert Myers, Krishnan Narayan, Yasunori Nomura, Masahiro Nozaki, Tokiro Numasawa, Shinsei Ryu, Tomonori Ugajin, Guifre Vidal and Aron Wall for useful conversations.
We are very much grateful to Pawel Caputa, Nilay Kundu and Masamichi Miyaji
for careful readings of the draft of this article and for valuable comments.
TT is supported by the Simons Foundation through the ``It from Qubit'' collaboration, by JSPS Grant-in-Aid for Scientific Research (A) No.16H02182 and by JSPS Grant-in-Aid for Challenging Research (Exploratory) 18K18766. TT is also supported by World Premier International Research Center Initiative (WPI Initiative) from the Japan Ministry of Education, Culture, Sports, Science and Technology (MEXT). TT is very grateful to the workshop ``Entanglement in Quantum Systems'' in GGI, Florence and the workshop ``New Frontiers in String Theory 2018'' in YITP, Kyoto (YITP-T-18-04), where parts of this work were conducted.

\begin{appendix}

\section{A Derivation of Liouville Action from AdS$_3$}
Consider a Euclidean Poincare AdS$_3$, given by the metric $ds^2=R_{AdS}^2 (dz^2+dT^2+dX^2)/z^2$,
and introduce a position dependent cut off defined by
\be
z\geq \ep\cdot  e^{-\ti{\phi}(T,X)}.  \label{cutpos}
\ee
If we set $\ti{\phi}=0$, then this is the usual UV cut off with the homogeneous lattice spacing given by $\ep$. Here we assume $\ti{\phi}$ is a non-trivial function of $T$ and $X$. As we will see later, $(T,X,\ti{\phi})$ are closely related to $(t,x,\phi)$ in the (\ref{cftm}), where the path-integral optimization for two dimensional CFTs was explained.

For the position dependent cut off, the metric on the boundary $M$ specified by
$z=\ep\cdot e^{-\ti{\phi}}$ reads
\be
ds^2=\frac{e^{2\ti{\phi}}}{\ep^2}\left[\left(1+\ep^2 e^{-2\ti{\phi}}(\de_T\ti{\phi})^2\right)dT^2
+2\ep^2(\de_T\ti{\phi})(\de_X\ti{\phi})dTdX+\left(1+\ep^2 e^{-2\ti{\phi}}(\de_X\ti{\phi})^2\right)dX^2\right].  \label{indmetp}
\ee
The extrinsic curvature $K$ on this boundary surface $M$ is found to be
\be
K=R_{AdS}^{-1}\cdot \left(2-\ep^2 e^{-2\ti{\phi}}(\de_T^2+\de_X^2)\ti{\phi}\right).
\ee

The bulk gravity action on this three dimensional spacetime $N$ can be evaluated as follows
\ba
I^E_G &=& \frac{1}{4\pi G_NR_{AdS}^2}\int_N\s{g}-\frac{1}{8\pi G_N}\int_M \s{\gamma}K \no
&=& \frac{R_{AdS}}{4\pi G_N}\int dTdX \int^\infty_{\ep e^{-\ti{\phi}}}\frac{dz}{z^3} \no
&&\ \ \ -\frac{R_{AdS}}{8\pi G_N}\int dT dX \frac{e^{2\ti{\phi}}}{\ep^2}\left(2-\ep^2 e^{-2\ti{\phi}}(\de_T^2+\de_X^2)\ti{\phi}\right)
\s{1+\ep^2e^{-2\ti{\phi}}\left((\de_T\ti{\phi})^2+(\de_X\ti{\phi})^2\right)}\no
&&=-\frac{c}{12\pi}\int dTdX \left[\frac{e^{2\ti{\phi}}}{\ep^2}+(\de_T\ti{\phi})^2+
(\de_X\ti{\phi})^2\right], \label{actionlp}
\ea
where we neglected surface terms.

In order to compare with the argument in section \ref{pio}, we need to adjust the boundary metric
into the form (\ref{cftm}) via a coordinate transformation $T=t+\zeta(t,x)$ and $X=x+\eta(t,x)$, where
$\zeta$ and $\eta$ are infinitesimally small functions of order $O(\ep^2)$. Let us denote the Jacobian
of the transformation from $(t,x)$ to $(T,X)$ by $J$ such that $dTdX=Jdtdx$. By simplify equating
(\ref{indmetp}) and (\ref{pio}), we find
\be
e^{2\phi}
=J\cdot \frac{e^{2\ti{\phi}}}{\ep^2}\cdot \s{1+\ep^2e^{-2\ti{\phi}}\left((\de_T\ti{\phi})^2+(\de_X\ti{\phi})^2\right)}.
\ee
Thus we find
\be
\int dTdX \frac{e^{2\ti{\phi}}}{\ep^2}
\simeq \int dtdx \left[e^{2\phi}-\frac{1}{2}\left((\de_t\phi)^2+(\de_x\phi)^2\right)\right].
\ee
Therefore we can rewrite the action (\ref{actionlp}) in terms of $(t,x,\phi)$ as follows
(we keep terms up to $O(1)$ in the limit $\ep\to 0$):
\ba
I^E_G=-\frac{c}{24\pi}\int dtdx \left[(\de_t\phi)^2+(\de_x\phi)^2+2e^{2\phi}\right]. \label{actionlpp}
\ea
In this way we managed to show that $-I^E_G$ coincides with the Liouville action (\ref{lvlacet}) as
expected.\footnote{Notice that the coefficient of the potential term $\int e^{2\phi}$ does depend on the holographic regularization scheme.}

\section{Path-Integrals on dS$_2$/H$_2$ and Slices in AdS$_3$}

Here we would like to examine explicit examples of path-integral circuits for a two dimensional de Sitter space dS$_2$ and hyperbolic space H$_2$ in the light of a connection between path-integrals and tensor networks, which was recently found in \cite{Path}. These spaces appear as
special codimension one slices in Euclidean/Lorentzian Poincare AdS$_3$ given by (\ref{msig}),(\ref{msiga}) and
(\ref{msigb}). Each of the metric of these three surfaces is given by
\ba
&& \mbox{H}_2\ \  \mbox{in}\ \  \mbox{H}_3 \to (\ref{msig}):\   ds^2=\frac{dt^2+\sin^2\ap\ dx^2}{\cos^2\ap\ t^2}, \no
&& \mbox{dS}_2\ \  \mbox{in}\ \  \mbox{AdS}_3 \to (\ref{msiga}):\ ds^2=\frac{-dt^2+\cosh^2\theta\ dx^2}{\sinh^2\theta\ t^2},   \no
&& \mbox{H}_2\ \ \mbox{in}\ \  \mbox{AdS}_3 \to (\ref{msigb}):\  ds^2=\frac{dt^2+\sinh^2\ti{\theta} \ dx^2}{\cosh^2\ti{\theta}\ t^2}.  \no
\ea
For example, in the null limit $\theta\to \infty$ or $\ti{\theta}\to \infty$ we find from the above metrics that the radii of dS$_2$ and H$_2$ shrink to zero.

\subsection{Path-Intgerals on dS$_2$}

Let us consider a path-integral on a two dimensional Lorentzian spacetime defined by the metric:
\ba
ds^2=-du^2+R(u)^2 dy^2. \label{metdsa}
\ea
It is useful to define $h(u)=\frac{\dot{R}(u)}{R(u)}$. If $h(u)$ is a non-zero constant, this spacetime coincides with a de Sitter space.

As introduced in \cite{Caputa:2017urj} and reviewed in section 2.2, our definition of UV cut off is such that the lattice spacing is $\ep$
with respect to the length measured by the above $ds^2$. Therefore it is useful to introduce another coordinate instead of $y$:
\be
\xi=R(u)y,
\ee
so that the lattice spacing in the coordinate $\xi$ is given by $\Delta\xi=\ep$, which corresponds to the original
``$|\Psi(u)\lb$ picture'' in cMERA \cite{cMERA,NRT}.
Then we can rewrite the metric (\ref{metdsa}) as follows:
\ba
ds^2=-(1-h(u)^2\xi^2)du^2-2h(u)\xi d\xi du+d\xi^2.
\ea
For simplicity, consider a massless scalar $\vp$ in this spacetime, defined by the action
\ba
S&=&\int du d\xi\s{-g}\left[-\frac{1}{2}g^{\mu\nu}\de_\mu\vp\de_\nu \vp\right]\no
&=&\int du d\xi\frac{1}{2}\left[\dot{\vp}^2+2h(u)\xi\dot{\vp}\de_\xi\vp-(1-h(u)^2\xi^2)(\de_\xi\vp)^2\right].
\ea
The evolution in the $u$ direction is described by
\be
P\exp\left(-i\int du H_u\right),  \label{qcpth}
\ee
with the Hamiltonian
given by
\be
H_u=\int d\xi \left[\frac{1}{2}(\pi^2+\de_\xi\vp^2)-h(u)\xi\pi\de_\xi\vp\right],
\ee
where $\pi=\dot{\vp}+h\xi\de_\xi\vp$ is the conjugate momentum. With the UV cut off $\Delta \xi=\ep$, this $u-$evolution defines the quantum circuit we are interested.

Therefore we can then we can express $H(u)$ as follows:
\ba
H_u=H_0+h(u)\cdot L', \label{hamu}
\ea
where $L'$ is the dilatation (or equally relativistic scale transformation), following the notation in \cite{cMERA},
and $H_0=\int d\xi\frac{1}{2}(\pi^2+\de_\xi\vp^2)$ is the standard Hamiltonian in the flat space $h(u)=0$.
Remember that the cMERA was originally defined by the quantum circuit
P$\exp\left(-i\int du L'\right)$ for the scale below the UV cut off \cite{cMERA}.
Even though here we employ the free scalar model as an exmaple of 2d CFT, the result (\ref{hamu}) should be true for any CFTs as it only involves the conformal symmetry following the arguments in \cite{Path}.

It is important to note that $L'$ acts locally on local operators with a conformal dimension $\Delta$
such that
\be
e^{-iuL'}O_\Delta(x)e^{iuL'}=e^{\Delta u}\cdot O_\Delta(e^ux).  \label{localt}
\ee
In this sense, the evolution by $L'$ is different from that by the Hamiltonian $H_0$, which gives a relativistic propagations of excitations.

\subsection{Relation to Codimension One Surfaces in AdS$_3$}

Now we would like to consider how the quantum circuit (\ref{qcpth}) corresponds to the bulk codimension one surface
$M_\Sigma$ as we have discussed in the main context of this article.
If we set
\be
t=-\ep\cdot \frac{\cosh\theta}{\sinh\theta}\cdot e^{-u\sinh\theta},\ \ \ h(u)=\sinh\theta,
\ee
then the metric of the de Sitter space (\ref{msiga}) agrees with (\ref{metdsa}).
Thus this de Sitter space in Lorentzian AdS$_3$ is
interpreted as the quantum circuit (\ref{qcpth}) with
\be
H(u)=H_0+\sinh\theta\cdot L'.
\ee
Note that in the AdS$_3$, the AdS boundary corresponds to $u=0$, while
 the IR point with $z=z_0$ does to $u=-\frac{1}{\sinh\theta}\log\left(\frac{z_0}{\ep}\right)<0$.
Thus we find that the dS$_2$ (\ref{msiga}) for the range $\ep \leq z \leq z_0$ corresponds to the quantum circuit:
\be
P\exp\left(-i\int^0_{-\frac{1}{\sinh\theta}\log\left(\frac{z_0}{\ep}\right)} du \left(H_0+\sinh\theta\cdot L'\right)\right).  \label{qcptha}
\ee
If we set $\theta=0$, this is indeed reduced to the ordinary evolution in the flat space and this is consistent with that $M_\Sigma$ coincides with the AdS boundary. On the other hand, when $\theta=\infty$ dual to the null limit of
$M_\Sigma$, the circuit only includes
the dilatation as in the cMERA circuit.

We can repeat the analysis of the previous subsection for path-integral on the H$_2$.
We find the hyperbolic space (\ref{msigb}) in Lorentzian AdS$_3$ corresponds to the circuit
\ba
P\exp\left(-\int^0_{-\frac{1}{\cosh\ti{\theta}}\log\left(\frac{z_0}{\ep}\right)} du \left(H_0+i\cosh\ti{\theta}\cdot L'\right)\right).  \label{qcpthb}
\ea
Again, in the null limit $\ti{\theta}=\infty$, we find the dilation dominates the circuit.

On the other hand, in the Euclidean AdS$_3$  (i.e. H$_3$), the surface (\ref{msig}) corresponds to the quantum circuit:
\ba
P\exp\left(-\int^0_{-\frac{1}{\cos\ap}\log\left(\frac{z_0}{\ep}\right)} du \left(H_0+i\cos\ap\cdot L'\right)\right).  \label{qcpthh}
\ea

In these examples,\footnote{Note also that all these results (\ref{qcptha}), (\ref{qcpthb}) and (\ref{qcpthh}) agree with the expectations in \cite{MTW} from the Killing symmetry of AdS$_3$, that the evolutions on both de-Sitter and hyperbolic slices are due to the dilation, via a suitable shift of the time coordinate $t\to t+t_0$ in the definition of the dilation.}
 it is intriguing to note that the amount of $H_0$ evolutions is proportional to the geodesic length $\int \s{g}$, while that of $L'$ is proportional to the growth of the entanglement entropy $S_A-S_{\ti{A}}$ compared the final state with the initial state. The former is the quantity
we considered in section 4. As argued in section 4, the length in AdS$_3$ (or the area in higher dimensional AdS) is expected to count the number of unitary gates which scramble quantum states.
Here we do not include the dilation $L'$ in such unitary gates as it does not scramble the quantum state as can be seen from the local transformation property (\ref{localt}). In the null limit, since we only have the dilatation gate $L'$, the length (i.e. the right hand side of (\ref{eee})) gets vanishing.

\end{appendix}


\end{document}